\pdfoutput=1
\documentclass[10pt,twocolumn,letterpaper]{article}
\usepackage[pagenumbers]{cvpr}

\usepackage{graphicx}	
\usepackage{amsmath}	
\usepackage{amssymb}	
\usepackage{placeins}
\usepackage{booktabs}
\usepackage{times}
\usepackage{microtype}
\usepackage{epsfig}
\usepackage{caption}
\usepackage{float}
\usepackage{placeins}
\usepackage{color, colortbl}
\usepackage{stfloats}
\usepackage{enumitem}
\usepackage{tabularx}
\usepackage{xstring}
\usepackage{multirow}
\usepackage{xspace}
\usepackage{url}
\usepackage{subcaption}
\usepackage{xcolor}
\newcolumntype{Y}{>{\centering\arraybackslash}X}
\definecolor{groupblue}{RGB}{36,91,150}
\definecolor{oursbg}{RGB}{241,241,241}

\usepackage[hang,flushmargin]{footmisc}

\newcommand{\supp}{appendix\xspace}

\newcommand{\R}[1]{{%
    \textbf{%
        \ifstrequal{#1}{1}{\textcolor{red}{R#1}}{%
        \ifstrequal{#1}{2}{\textcolor{blue}{R#1}}{%
        \ifstrequal{#1}{3}{\textcolor{magenta}{R#1}}{%
        \ifstrequal{#1}{4}{\textcolor{teal}{R#1}}{%
                           \textcolor{cyan}{R#1}%
        }}}}%
    }%
}}

\usepackage{tikz}
\usetikzlibrary{arrows.meta,decorations.pathreplacing,positioning}
\tikzset{
  fwbox/.style={draw=#1!65!black, fill=#1!8, rounded corners=3pt, minimum height=8mm, inner sep=4pt, align=center},
  fwbox/.default=gray,
  fwlab/.style={font=\sffamily\scriptsize, text=black!65, align=center},
  fwhead/.style={font=\sffamily\bfseries, anchor=west},
  fwflow/.style={->, thick, draw=black!70},
  fwtgt/.style={circle, fill=black!75, inner sep=1.15pt},
  fwpop/.style={circle, fill=blue!70!black, inner sep=1.15pt},
  fwimg/.style={circle, fill=orange!85!black, inner sep=1.15pt},
}

\definecolor{cvprblue}{rgb}{0.21,0.49,0.74}
\usepackage[pagebackref,breaklinks,colorlinks,allcolors=cvprblue]{hyperref}
\usepackage[capitalize]{cleveref}
\crefname{section}{Sec.}{Secs.}
\crefname{table}{Table}{Tables}
\crefname{figure}{Fig.}{Figs.}
\crefname{appendix}{App.}{Apps.}
\newcommand{\chorusAuthors}{%
  Jiayu Zuo\textsuperscript{1,2}\quad
  Dongyang Li\textsuperscript{2}\quad
  Chen Wei\textsuperscript{2,*}\quad
  Mo Wang\textsuperscript{1,2,*}\quad
  Quanying Liu\textsuperscript{1,2,*}\\[4pt]
  {\normalsize\textsuperscript{1}Southern University of Science and Technology}\\
  {\normalsize\textsuperscript{2}Omni-Intelligence}\\[2pt]
  {\small\textsuperscript{*}Co-corresponding authors}%
}
\newcommand{\chorusAuthorMetadata}{Jiayu Zuo, Dongyang Li, Chen Wei, Mo Wang, Quanying Liu}

\hypersetup{pdftitle={CHORUS: Cross-Subject Harmonization via Population Relay for Unseen Subjects in Brain Visual Decoding},pdfauthor={\chorusAuthorMetadata}}
\expandafter\def\csname KshotBHA10Mean\endcsname{41.57}
\expandafter\def\csname KshotBHA10SD\endcsname{0.70}
\expandafter\def\csname DirectKshotBHA10Mean\endcsname{36.83}
\expandafter\def\csname DirectKshotBHA10SD\endcsname{0.53}
\expandafter\def\csname DeltaKshotBHA10\endcsname{4.74}
\expandafter\def\csname KshotBHA20Mean\endcsname{45.23}
\expandafter\def\csname KshotBHA20SD\endcsname{0.68}
\expandafter\def\csname DirectKshotBHA20Mean\endcsname{39.84}
\expandafter\def\csname DirectKshotBHA20SD\endcsname{0.58}
\expandafter\def\csname DeltaKshotBHA20\endcsname{5.39}
\expandafter\def\csname KshotBHA50Mean\endcsname{51.36}
\expandafter\def\csname KshotBHA50SD\endcsname{0.65}
\expandafter\def\csname DirectKshotBHA50Mean\endcsname{45.26}
\expandafter\def\csname DirectKshotBHA50SD\endcsname{0.77}
\expandafter\def\csname DeltaKshotBHA50\endcsname{6.10}
\expandafter\def\csname KshotBHA100Mean\endcsname{56.06}
\expandafter\def\csname KshotBHA100SD\endcsname{0.92}
\expandafter\def\csname DirectKshotBHA100Mean\endcsname{49.13}
\expandafter\def\csname DirectKshotBHA100SD\endcsname{0.79}
\expandafter\def\csname DeltaKshotBHA100\endcsname{6.93}
\expandafter\def\csname KshotBHA200Mean\endcsname{60.41}
\expandafter\def\csname KshotBHA200SD\endcsname{0.73}
\expandafter\def\csname DirectKshotBHA200Mean\endcsname{54.08}
\expandafter\def\csname DirectKshotBHA200SD\endcsname{0.99}
\expandafter\def\csname DeltaKshotBHA200\endcsname{6.33}
\expandafter\def\csname KshotSAMGA10Mean\endcsname{41.34}
\expandafter\def\csname KshotSAMGA10SD\endcsname{0.65}
\expandafter\def\csname DirectKshotSAMGA10Mean\endcsname{37.14}
\expandafter\def\csname DirectKshotSAMGA10SD\endcsname{0.10}
\expandafter\def\csname DeltaKshotSAMGA10\endcsname{4.20}
\expandafter\def\csname KshotSAMGA20Mean\endcsname{44.44}
\expandafter\def\csname KshotSAMGA20SD\endcsname{0.60}
\expandafter\def\csname DirectKshotSAMGA20Mean\endcsname{39.96}
\expandafter\def\csname DirectKshotSAMGA20SD\endcsname{0.18}
\expandafter\def\csname DeltaKshotSAMGA20\endcsname{4.48}
\expandafter\def\csname KshotSAMGA50Mean\endcsname{49.54}
\expandafter\def\csname KshotSAMGA50SD\endcsname{0.71}
\expandafter\def\csname DirectKshotSAMGA50Mean\endcsname{44.79}
\expandafter\def\csname DirectKshotSAMGA50SD\endcsname{0.50}
\expandafter\def\csname DeltaKshotSAMGA50\endcsname{4.75}
\expandafter\def\csname KshotSAMGA100Mean\endcsname{53.48}
\expandafter\def\csname KshotSAMGA100SD\endcsname{0.33}
\expandafter\def\csname DirectKshotSAMGA100Mean\endcsname{48.92}
\expandafter\def\csname DirectKshotSAMGA100SD\endcsname{0.26}
\expandafter\def\csname DeltaKshotSAMGA100\endcsname{4.56}
\expandafter\def\csname KshotSAMGA200Mean\endcsname{57.43}
\expandafter\def\csname KshotSAMGA200SD\endcsname{0.24}
\expandafter\def\csname DirectKshotSAMGA200Mean\endcsname{52.81}
\expandafter\def\csname DirectKshotSAMGA200SD\endcsname{0.16}
\expandafter\def\csname DeltaKshotSAMGA200\endcsname{4.62}

\newfloat{algorithm}{tbp}{loa}
\floatname{algorithm}{Algorithm}
\crefname{algorithm}{Algorithm}{Algorithms}
\begin{document}
\title{CHORUS: Cross-Subject Harmonization via Population Relay\\for Unseen Subjects in Brain Visual Decoding}
\author{\chorusAuthors}
\maketitle
\begin{abstract}
EEG-based image retrieval performs well on training subjects but generalizes poorly to unseen individuals.
Existing calibration methods address this gap by aligning a new subject directly with image embeddings.
We find that the training population is a better alignment target: an unseen subject's embeddings are closer to the stimulus-matched population average than to the image embeddings, and the remaining gap to the population is largely a shift and a rotation.
Motivated by this finding, we propose CHORUS.
At training, stimulus-matched cross-subject averaging (SCA) teaches the encoder to match responses to the same image across subjects; at deployment, a two-stage calibration first aligns an unseen subject to population anchors predicted from the decoder's visual features, and then to the image embeddings.
On leave-one-subject-out THINGS-EEG2 with two backbones, CHORUS improves over direct calibration by 10.92 and 6.63 points on unlabeled test batches and by 4.2--6.9 points with 10--200 labeled pairs; even without the population stage, SCA and direct calibration each outperform prior training and calibration schemes.
The complete system reaches 78.37\% and 73.40\% 200-way Top-1 accuracy, 36.32 and 20.93 points above SCORE.
\end{abstract}

\section{Introduction}
\label{sec:intro}
\begin{figure}[t]
\centering
\includegraphics[width=\columnwidth]{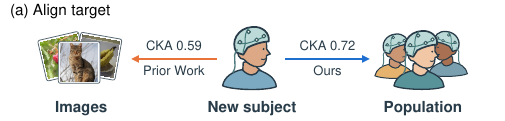}\par\nobreak
\includegraphics[width=\columnwidth]{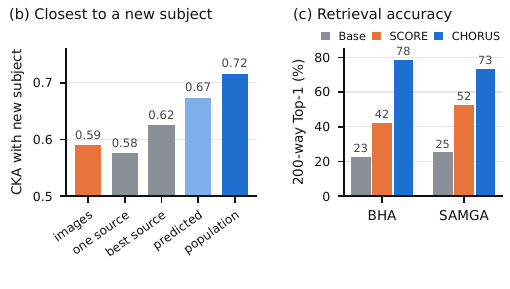}
\caption{\textbf{Where should a new subject be aligned?} (a) Across 30 folds, mean pre-calibration linear CKA over 200 stimuli is 0.72 with measured population EEG and 0.59 with image embeddings. (b) Mean CKA for all five references (BHA + SCA, best source selected post hoc). (c) Unlabeled-batch retrieval Top-1.}
\label{fig:hook}
\end{figure}

Visual decoding asks what a person is looking at given a recording of their brain activity. Decoders can now reconstruct natural scenes from functional MRI~\cite{takagi2023high,chen2023seeing,ozcelik2023natural,scotti2023reconstructing} and retrieve images from electroencephalography (EEG) or magnetoencephalography (MEG)~\cite{song2024decoding,li2024visual,benchetrit2024brain}. In EEG-based image retrieval, the decoder must identify which of several candidate images a person saw. A decoder trained and tested on the same person can learn how that individual responds to different images. Using it with someone new requires recognizing the same visual content across two people's recordings. How much can a decoder learn about that correspondence from the people it has already seen?

EEG-to-image retrieval methods typically train an encoder to align EEG responses with embeddings from a pretrained vision model such as CLIP~\cite{radford2021learning,song2024decoding,li2024visual}. For a new subject, collecting labeled EEG--image pairs is costly, so recent methods calibrate the decoder using unlabeled responses. SATTC~\cite{huang2026sattc} adjusts EEG--image similarity scores using the neighborhoods and ranks of candidate matches. SCORE~\cite{cui2026score} selects tentative EEG--image pairs and uses them to rotate the subject's EEG embeddings toward the corresponding image embeddings. Both take the candidate images as the reference from the outset, when the correspondence between responses and images is still unknown.

But is the image space the best calibration target? \Cref{fig:hook}(a) illustrates two possible references for a new subject: the training population's mean EEG response to each stimulus and the corresponding image embedding. We compared both with the held-out subject before calibration using linear centered kernel alignment (CKA)~\cite{kornblith2019similarity} across the 200 test stimuli in THINGS-EEG2~\cite{gifford2022large}. Across 30 folds, mean CKA was 0.72 for EEG averaged across nine training subjects, 0.59 for image embeddings, and 0.62 even for the best individual training subject (\cref{fig:hook}(b)). For a given image, the population and the new subject tend to deviate from its visual embedding in the same direction. A shift and a rotation then account for much of the remaining difference between the subject and the population (\cref{sec:analyses}).

Motivated by this finding, we introduce CHORUS, which calibrates a new subject against the population response predicted for each candidate image before matching the subject's EEG to the images.

Our main contributions are:

\noindent\textbf{1. The training population provides a closer initial reference for a new subject than image embeddings do.} Averaging preserves the image-specific direction in which different subjects' EEG embeddings tend to deviate from the visual embedding.

\begin{figure*}[t]
\centering
\includegraphics[width=0.96\textwidth]{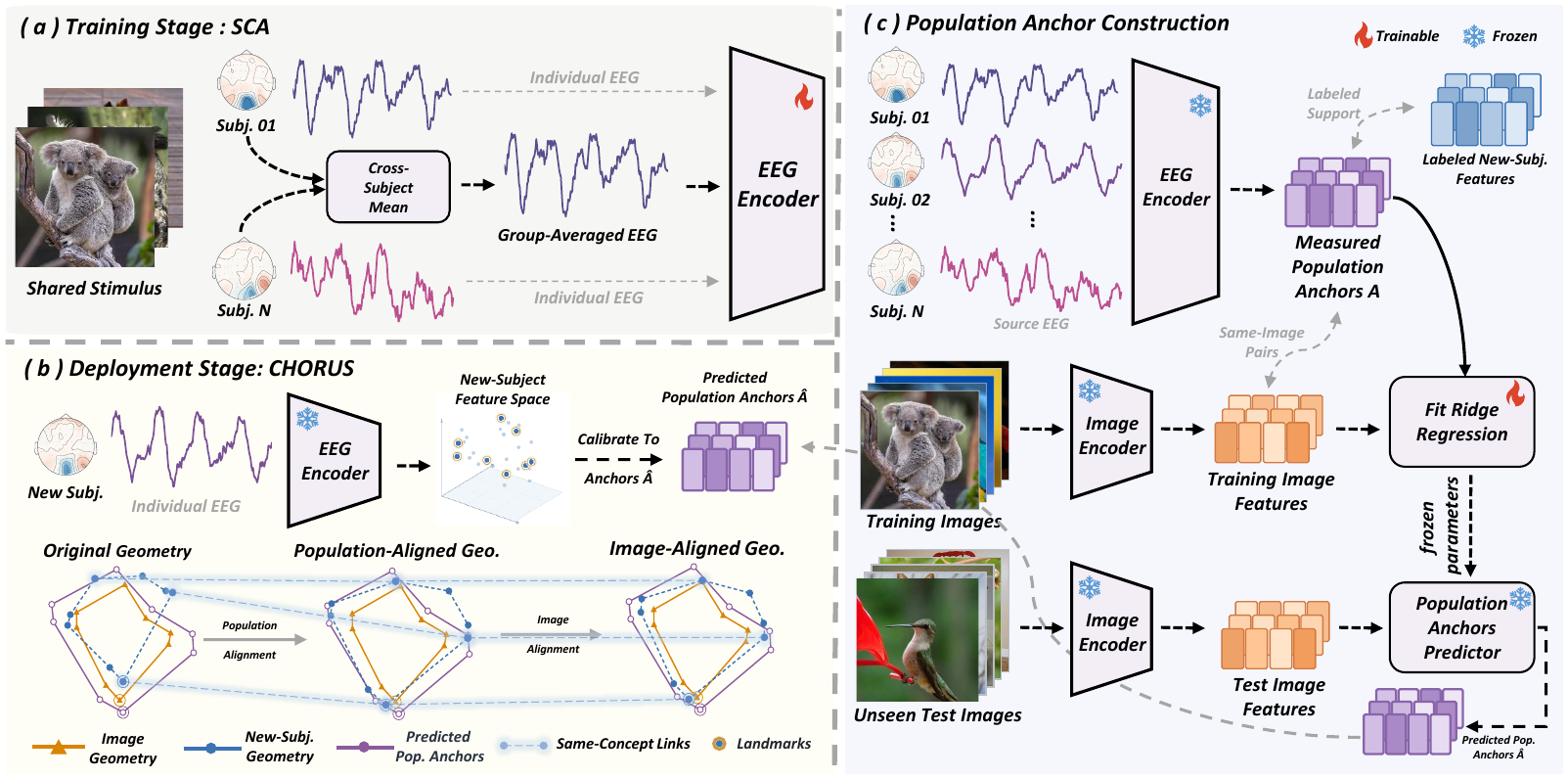}
\caption{\textbf{CHORUS.} (a) Training: SCA averages same-image EEG across subjects. (b) Deployment: target embeddings align first to population anchors, then to images; only the final representation is scored. (c) Measured anchors average source responses; predicted anchors cover unseen images.}
\label{fig:framework}
\end{figure*}

{\clubpenalty=10000\relax
\noindent\textbf{2. CHORUS learns shared stimulus responses during training and uses them to guide calibration.} Stimulus-matched cross-subject averaging (SCA) trains the encoder on EEG waveforms averaged across people viewing the same image. For an unlabeled batch, CHORUS first aligns the new subject to population anchors predicted from the candidate images, then refines the EEG--image matches (\cref{fig:framework}).\par}

\noindent\textbf{3. Population alignment improves cross-subject visual decoding, and the learned population map transfers to new images.} Compared with direct calibration, CHORUS gains 10.92 points on BHA~\cite{anon2027bha} and 6.63 on SAMGA~\cite{samga2026} in unlabeled 200-way retrieval and improves all six feature-based reconstruction metrics on BHA. When the same subject views new images, reusing the population map and refitting only the image stage comes within one point of refitting both stages on each backbone.

\section{Related Work}
\label{sec:related}

\paragraph{Visual decoding from brain activity.}
EEG studies first classified object categories~\cite{simanova2010identifying,kaneshiro2015representational,spampinato2017deep}, learned features end to end~\cite{schirrmeister2017deep}, and generated images~\cite{kavasidis2017brain2image}. From fMRI, multimodal models decoded visual categories~\cite{du2023decoding}, and diffusion models reconstructed images~\cite{takagi2023high,chen2023seeing,ozcelik2023natural,scotti2023reconstructing}.
On shared-stimulus datasets~\cite{gifford2022large,grootswagers2022human,hebart2023things}, EEG and MEG encoders aligned with vision models enable zero-shot retrieval and reconstruction~\cite{song2024decoding,li2024visual,benchetrit2024brain,zhang2025cognitioncapturer,li2025brainflora,wu2025bridging,zheng2026learning,zhang2026neurobridge,jo2026hyfi,anon2027bha,dai2026autoregressive}, mostly within subjects; we target unseen subjects.

\paragraph{Cross-subject transfer.}
Hyperalignment~\cite{haxby2011common}, the shared response model~\cite{chen2015reduced}, generalized Procrustes analysis~\cite{gower1975generalized}, and optimal-transport alignment~\cite{thual2022aligning} map subjects to a common space; fMRI decoders align a new subject to a reference subject or a shared latent space~\cite{ho2023inter,ferrante2024through,wang2024mindbridge,quan2024psychometry,scotti2024mindeye2,xia2024umbrae}. In the reverse direction, shared encoding models predict cortical responses to images~\cite{yu2025braincorl}. EEG methods re-center or rotate covariances per subject~\cite{he2020transfer,rodrigues2019riemannian}, and inter-subject correlation shows that stimulus-driven responses are shared~\cite{dmochowski2012correlated}.
For EEG-to-image decoding, SAMGA builds subject-aware visual targets~\cite{samga2026}, SCORE simulates coordinate recovery during training~\cite{cui2026score}, and ProCA adapts semantic supervision~\cite{zhou2026proca}.
CHORUS keeps the decoder frozen, uses the source average only as an intermediate calibration target that it predicts for unseen stimuli, and needs no target labels; SCA is a cross-subject form of input mixing~\cite{zhang2018mixup}.

\paragraph{Test-time calibration.}
SATTC combines subject-adaptive whitening, similarity scaling, and structural ranking~\cite{huang2026sattc}; SCORE recovers an orthogonal alignment from unlabeled EEG--image landmarks~\cite{cui2026score}.
Both target image space. CORAL aligns domain covariances~\cite{sun2016return}, optimal transport matches domain distributions~\cite{courty2017optimal}, and querybank normalization adjusts retrieval scores~\cite{bogolin2022cross}. Other test-time methods update model parameters~\cite{wang2021tent,liang2020shot,li2024ttime}.
CHORUS separates the deviation shared across people from the subject-specific gap and reaches the images through the population. Its correspondence inference adapts Gromov--Wasserstein coupling~\cite{alvarez2018gromov,vayer2019optimal} and matching--Procrustes alternation~\cite{schonemann1966generalized,artetxe2018robust,grave2019unsupervised} with Hungarian assignment~\cite{kuhn1955hungarian} and CSLS~\cite{conneau2018word}.
NEAR predicts image-specific neural anchors from a high-repetition center~\cite{cui2026near}; we predict the population's average response. Independently trained models share enough geometry to align with few pairs~\cite{moschella2023relative,huh2024platonic,schnaus2026rosetta}; a decoder's embeddings of different people share more geometry with each other than with the images.

\section{CHORUS}
\label{sec:method}

CHORUS has a training part, SCA, and a deployment part that calibrates the frozen decoder in two stages, from the new subject to population anchors and from there to the images (\cref{fig:framework}).
\Cref{sec:sca} trains the encoder so that a new subject's responses can be matched to the population, \cref{sec:anchors} defines the population anchors and explains why they are the right intermediate target, and \cref{sec:gacr,sec:deployment} describe the two-stage calibrator and how it adapts to each deployment budget.

\subsection{Setup}
\label{sec:setup}
Let $\mathcal S$ be the training (source) subjects and $u\notin\mathcal S$ a new target subject.
An encoder $f$ maps a preprocessed EEG response $x_{p,i}$ of subject $p$ to image $I_i$ to an embedding $h_{p,i}=f(x_{p,i})$, and a frozen image branch gives the image embedding $v_i$.
Retrieval ranks candidate images by the backbone's score $s_0(h,I_j)$, a cosine similarity in our backbones.
The target's responses are never used for training; they enter only the calibration permitted by the deployment budget.

\subsection{Learning the shared response}
\label{sec:sca}
Calibrating a new subject to the population requires telling which population response each of the new subject's responses belongs to; SCA trains the encoder for this cross-person correspondence.
For a group $G\subseteq\mathcal S$ of subjects who viewed image $I_i$, SCA forms the averaged waveform $\widetilde x_{G,i}=\frac{1}{|G|}\sum_{p\in G}x_{p,i}$\label{eq:sca} and adds the pair $(\widetilde x_{G,i},I_i)$ to the training data, alongside the individual recordings.
Averaging suppresses subject-specific noise in the input, so the encoder learns which image a response belongs to across people; this sharpens the per-image correspondence that calibration relies on (\cref{sec:gain}), at no cost in losses or parameters.
For BHA and SAMGA we keep every individual recording in each epoch. For each such recording we draw other source subjects who saw the same image and average their waveforms in groups of two or four. Individual, pair-averaged and four-subject examples occur in a $4{:}6{:}1$ ratio. Resampling the donors lets the same image appear with different combinations of subjects while retaining its original training label.

\subsection{Population anchors}
\label{sec:anchors}
\begin{figure}[!t]
\centering
\includegraphics[width=\columnwidth]{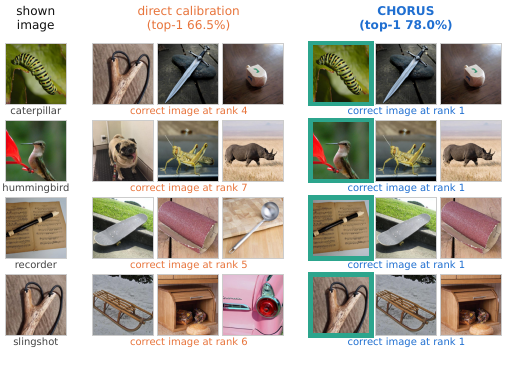}
\caption{\textbf{Direct calibration versus CHORUS.} Top-three predictions for the held-out subject closest to the mean CHORUS accuracy (BHA + SCA); green: correct. Shown: four random queries that only CHORUS gets right and direct calibration ranks outside its top three.}
\label{fig:qual}
\end{figure}

After training, the \emph{population anchor} of image $I_i$ is the average of the source subjects' encoded responses,
\begin{equation}
a_i=\frac{1}{|\mathcal S|}\sum_{p\in\mathcal S}f(x_{p,i}).
\label{eq:population_anchor}
\end{equation}
Measured anchors are available for images the source subjects saw.
For unseen candidate images, let $\widehat A=\{\widehat a_j\}$ denote their predicted population anchors, one per image.
SCA averages waveforms before encoding. The training targets for the anchor predictor come from \cref{eq:population_anchor}: it encodes each source subject separately and averages the embeddings for each training image.
Let $m_i$ contain the pooled features of the visual layers used by the decoder, $v_i$ its final image embedding and $p_i$ their patch-token PCA features. The residual predictor has the form
\[
\widehat a_i=g_{\rm pool}(m_i,v_i)+g_{\rm patch}(p_i).
\]
The first ridge regression predicts the measured anchor from global features. Pooling loses spatial information, so the second ridge regression uses patch tokens to predict the part missed by the first. For each training image, the residual target is its measured anchor minus the pooled prediction obtained when that image was held out of the regression fit. We fit the patch projection on source-training images and choose the predictor on source validation by centered cosine with measured anchors (\supp).
The anchor set is fixed before calibration: it uses neither the target's labels nor source EEG responses to the candidate test images.
\Cref{fig:qual} illustrates how this choice of reference can change which image ranks first.

\paragraph{Why a population reference.}
After rigid alignment, model subject $p$'s embedding of image $i$ as $h_{p,i}=v_i+\delta_i+e_{p,i}$: the image embedding, a deviation $\delta_i$ shared across people, and an independent subject-specific term of power $\eta^2$. Then (derivation in the \supp)
\begin{equation}
\mathbb E\|h_{u,i}-y\|^2=
\begin{cases}
\eta^2+\|\delta_i\|^2, & y=v_i\ \text{(image)},\\
2\eta^2, & y=h_{p,i}\ \text{(one source)},\\
\eta^2+\eta^2/|\mathcal S|, & y=a_i\ \text{(population)}.
\end{cases}
\label{eq:reference_error}
\end{equation}
Averaging cancels the idiosyncratic terms but keeps the shared deviation that the image lacks.
Under this model, the population is therefore closer in expectation to a new subject than a single source, and closer than the image when $\|\delta_i\|^2>\eta^2/|\mathcal S|$, whereas a single source needs $\|\delta_i\|^2>\eta^2$.
The data follow this ordering (\cref{fig:hook}(b), \cref{sec:analyses}), and predicted anchors, which approximate $v_i+\delta_i$, retain part of the population's advantage. The model ignores correlations between subjects, so it predicts orderings, not magnitudes.

The model also separates the two stages: the new subject--population gap is subject-specific, whereas the population--image gap $\delta_i$ is common to everyone. CHORUS therefore fits a per-subject shift and rotation in the population stage and treats the image stage as a property of the decoder: the anchor predictor with labeled support, and an image-space calibration in unlabeled batches.

\begin{table}[!tb]
\centering\small\setlength{\tabcolsep}{2pt}
\renewcommand{\arraystretch}{1.15}
\caption{\textbf{Training schemes without calibration.} 200-way Top-1 (\%) of each backbone trained with each objective; means over three seeds and ten held-out subjects. With calibration: \supp.}
\label{tab:training}\label{tab:ablation}
\begin{tabularx}{\columnwidth}{@{}l*{4}{Y}>{\columncolor{oursbg}\hspace{3pt}}c<{\hspace{3pt}}@{}}
\toprule
Backbone & Base & SATTC & SCORE & ProCA & SCA (ours)\\
\midrule
BHA & 22.67 & 23.43 & 21.27 & 24.57 & \textbf{28.17}\\
SAMGA & 25.22 & 25.48 & 28.47 & 25.87 & \textbf{36.62}\\
\bottomrule
\end{tabularx}
\end{table}

\subsection{Population-relay calibration}
\label{sec:gacr}
Calibration keeps the encoder and scorer frozen. For an unlabeled batch, the population stage shifts the target EEG mean to the source training-population mean and initializes its transform from transport landmarks~\cite{vayer2019optimal}. The transport cost combines the cosine distance between a query and an anchor with agreement between their respective within-set distances. Thus a tentative pair is supported when other queries near it also have nearby partners in the anchor set. We select pairs that are mutual maxima of the transport coupling and weight them by their coupling mass in the initial fit. In each subsequent matching step, let $q_i^{(s)}$ be the current EEG embedding and $r_j^{(s)}$ the reference for candidate image $j$ in stage $s$. Hungarian matching~\cite{kuhn1955hungarian} maximizes total cosine similarity over $\mathcal P$, the set of assignments mapping the $n$ queries to distinct candidates:
\begin{equation}
\pi_s^\star=\mathop{\arg\max}_{\pi\in\mathcal P}
\sum_{i=1}^{n}\cos\!\left(q_i^{(s)},r_{\pi(i)}^{(s)}\right).
\label{eq:stage_assignment}
\end{equation}
The population stage matches to predicted anchors ($r_j^{(\mathrm{pop})}=\widehat a_j$). The image stage starts from the population-aligned EEG and solves \cref{eq:stage_assignment} anew with image embeddings ($r_j^{(\mathrm{img})}=v_j$), so a query can change partners.

For the population stage of unlabeled calibration, given matched pairs $(z_k,y_k)$ of target embeddings and references, and normalized nonnegative weights $w_k$, we fit a regularized orthogonal recovery,
\begin{equation}
R^\star=\mathop{\arg\min}_{R^\top R=I}\sum_k w_k\|(z_k-\mu_z)R-(y_k-\mu_y)\|^2+\rho\,\sigma\|R-I\|_F^2,
\label{eq:recovery}
\end{equation}
and map an embedding $h$ to $T(h)=(h-\mu_z)R^\star+\mu_y$, where $\mu_z,\mu_y$ are weighted pair means, $\sigma$ is the largest singular value of the weighted cross-covariance, and $\rho$ keeps $R^\star$ close to the identity ($\rho=0.1$).
Its closed-form solution is a polar factor~\cite{schonemann1966generalized} (\supp), so $T$ is a rigid motion: it cannot distort the subject's internal geometry.
In unlabeled calibration, we restrict this map to a shift and a rotation because this family captures most of the gap between a new subject and the population, whereas the tested more flexible maps perform worse (\cref{sec:analyses}).

The discrete pairs then receive equal weight. In the population stage we alternate assignment and recovery until the assignment repeats or the iteration limit is reached. The current transform is used to find the next pairs, but each new transform is fitted to the fixed input of its stage: mean-shifted EEG in the first stage and population-aligned EEG in the second. In the image stage, recovery first matches per-dimension means and standard deviations on the inferred pairs, then fits the rotation in \cref{eq:recovery}; the image transform thus also includes diagonal scaling.
Retrieval uses only the resulting image-aligned embedding, with CSLS~\cite{conneau2018word}:
\begin{equation}
\widehat j_i=\arg\max_j\operatorname{CSLS}\!\left(T_{\rm image}(T_{\rm pop}(h_i)),v_j\right).
\label{eq:final_score}
\end{equation}
The one-to-one constraint in \cref{eq:stage_assignment} applies only while fitting the transforms. At readout, each EEG query ranks the gallery independently, so two queries can return the same image.
Its building blocks (orthogonal recovery, transport, Hungarian matching, CSLS) are standard; the calibrator built from them is new: it aligns the new subject to a predicted population rather than to the images, infers correspondences anew in two successive spaces, refits every update from the fixed input of its stage, and, with labeled support, lets the anchor predictor take the place of the image stage (\cref{sec:deployment}).
\emph{Direct calibration} shifts toward the gallery mean, initializes correspondences with transport, and performs iterative recovery directly against the image embeddings. It shares the encoder, gallery, and final CSLS readout with CHORUS, so comparing the two isolates the population stage.

\begin{table}[!tb]
\centering
\caption{\textbf{Calibration schemes and complete systems.} Unlabeled 200-way Top-1 (\%); means over three seeds and ten held-out subjects (streams: also three arrival orders).}
\label{tab:retrieval}\label{tab:calib}\label{tab:system}
\begingroup
\small
\setlength{\tabcolsep}{2pt}
\renewcommand{\arraystretch}{1.15}
\newcommand{\blk}[2]{\multicolumn{#1}{@{}l}{\footnotesize\textit{#2}}}
\begin{tabularx}{\columnwidth}{@{}l*{4}{Y}@{}}
\toprule
& \multicolumn{2}{c}{BHA} & \multicolumn{2}{c}{SAMGA}\\
\cmidrule(lr){2-3}\cmidrule(l){4-5}
Method & Batch & Stream & Batch & Stream\\
\midrule
\blk{5}{(a) Calibration of the same SCA encoder}\\
SATTC calib.~\cite{huang2026sattc} & 40.15 & 32.68 & 36.87 & 29.20\\
SCORE calib.~\cite{cui2026score} & 56.48 & 46.57 & 56.52 & 43.87\\
Direct calib. & 67.45 & 53.63 & 66.77 & 50.93\\
\rowcolor{oursbg}
Population relay (ours) & \textbf{78.37} & \textbf{63.13} & \textbf{73.40} & \textbf{56.36}\\
\midrule
\blk{5}{(b) Complete systems, own training and calibration}\\
SATTC~\cite{huang2026sattc} & 35.23 & 29.03 & 29.85 & 24.11\\
SCORE~\cite{cui2026score} & 42.05 & 35.46 & 52.47 & 41.34\\
\rowcolor{oursbg}
CHORUS (SCA + relay) & \textbf{78.37} & \textbf{63.13} & \textbf{73.40} & \textbf{56.36}\\
\bottomrule
\end{tabularx}
\endgroup

\end{table}

\subsection{One relay for three deployment budgets}
\label{sec:deployment}
\textbf{Labeled support.} $K$ pairs of target EEG and the measured population anchor of the same training image fit only the population stage, $T_{{\rm pop},K}(h)=s\,(h-\bar h)R_u+\bar a$, with support means $\bar h,\bar a$, a global scale $s$ and $R_u$ from \cref{eq:recovery} at the fixed $\rho=10/K$.
Each test query is then scored on its own against the predicted anchors with CSLS, so the anchor predictor carries the population--image step; direct calibration fits the same map to the supports' images.
\textbf{Unlabeled batch.} Both stages infer correspondences jointly over test responses that arrive together.
\textbf{Causal streaming.} Trials arrive one at a time and the relay is refitted periodically from earlier trials only (\supp).

\section{Experiments}
\label{sec:experiments}

\subsection{Setup}
\label{sec:implementation}
\label{sec:retrieval_protocol}
We use THINGS-EEG2~\cite{gifford2022large} (ten subjects, 1{,}654 training and 200 test concepts) with leave-one-subject-out evaluation: each subject in turn is the unseen target and the other nine are sources.
Our main backbones are BHA~\cite{anon2027bha} and SAMGA~\cite{samga2026}, trained with their official code; \emph{Base} is each backbone without SCA or calibration.
A new subject brings one of three calibration budgets: $K\in\{10,20,50,100,200\}$ labeled pairs from its training-image recordings, an unlabeled batch of its 200 test responses, or a causal stream of test trials.
We compare SATTC~\cite{huang2026sattc} and SCORE~\cite{cui2026score} as complete systems and as calibrators on the same SCA encoder. The training comparison also includes ProCA~\cite{zhou2026proca}. At every budget, direct calibration (\cref{sec:gacr,sec:deployment}) shares everything with CHORUS except the population stage.
Constants that our calibrator shares with SCORE keep SCORE's values, so differences do not come from tuning.
We report 200-way Top-1, averaged over three training seeds and ten held-out subjects; statistical tests and protocol details are in the \supp. BHA distills visual teachers into a brain encoder. CHORUS calibrates frozen encoders for new subjects, including SAMGA. Code and checkpoints will be released.

\begin{figure}[!t]
\centering
\includegraphics[width=\columnwidth]{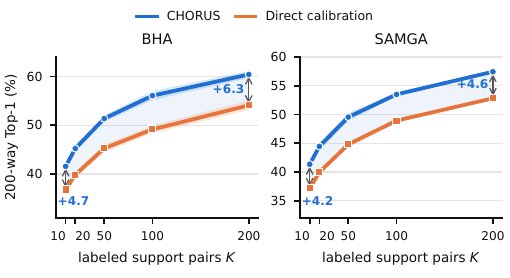}
\caption{\textbf{Accuracy with $K$ labeled pairs} (SCA encoder). Both routes fit the same map on the same supports; CHORUS targets measured anchors and scores predicted ones, direct calibration targets and scores images. Each query is scored on its own. Bands: seed SD; numbers: CHORUS lead.}
\label{fig:kcurve}
\end{figure}

\subsection{Main results}
\label{sec:retrieval_results}
\textbf{Training.} Before calibration, SCA improves retrieval over the next-best training scheme by 3.60 points on BHA and 8.15 on SAMGA (\cref{tab:training}).
\textbf{Calibration.} The choice of reference still matters after cross-subject training. With the same SCA encoder fixed, aligning first to predicted population anchors raises unlabeled-batch Top-1 from 67.45 to 78.37\% on BHA and from 66.77 to 73.40\% on SAMGA, improving retrieval for every held-out subject (\cref{tab:retrieval}(a)). For the BHA subject closest to mean CHORUS accuracy, the relay corrects 24 direct-calibration errors and introduces one (\cref{fig:qual}).
\textbf{Complete systems.} CHORUS exceeds SCORE by 36.32 and 20.93 points on BHA and SAMGA (\cref{tab:retrieval}(b)).

\textbf{With labeled support.} Does the population still help when the EEG--image pairs are known? Both routes fit the same type of map on the same $K$ labeled responses. Fitting to measured population anchors improves retrieval over fitting to image embeddings by 4--7 points at every budget on both backbones (\cref{fig:kcurve}). Even with correct pairings, the target labels are better spent estimating how the new subject differs from the population. The predictor learned from source subjects supplies the anchor for each candidate image.

\begin{figure}[!t]
\centering
\includegraphics[width=\columnwidth]{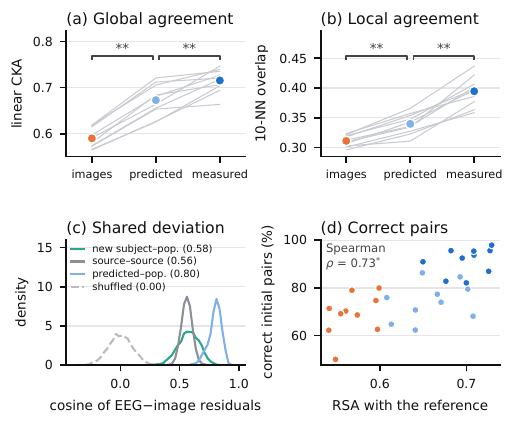}
\caption{\textbf{Population geometry} (BHA + SCA). (a,b) Global (linear CKA) and local (10-NN overlap) agreement of the held-out subject with images, predicted anchors and the measured population; gray: subjects. (c) Cosines of image-subtracted residuals. (d) Per subject, RSA with each reference versus correct initial transport pairs.}
\label{fig:geo}
\end{figure}

\textbf{Causal streaming.} When each trial is predicted before it enters the history, the relay outperforms direct calibration with the same history by 9.50 points on BHA and 5.43 on SAMGA, improving retrieval for all ten subjects on both. The population reference thus helps even before the full batch is available.

\begin{table}[!htb]
\centering\small\setlength{\tabcolsep}{2pt}
\renewcommand{\arraystretch}{1.15}
\caption{\textbf{Alljoined-1.6M.} Unlabeled-batch 200-way Top-1 (\%); one training seed, mean over 20 held-out subjects. SCORE: its calibration~\cite{cui2026score} of the same encoder.}
\label{tab:xdataset}
\begin{tabularx}{\columnwidth}{@{}l*{3}{Y}>{\columncolor{oursbg}\hspace{3pt}}c<{\hspace{3pt}}@{}}
\toprule
Encoder & No calib. & SCORE & Direct & Relay (ours)\\
\midrule
Base & 6.53 & 12.85 & 15.48 & \textbf{16.88}\\
SCA & 10.50 & 15.20 & 18.20 & \textbf{20.23}\\
\bottomrule
\end{tabularx}
\end{table}

\textbf{A second dataset.} On Alljoined-1.6M~\cite{alljoined2025} (consumer-grade 32-channel EEG), the batch advantage over direct calibration is smaller but persists with both encoders, at 1.40 points with Base and 2.03 with SCA (\cref{tab:xdataset}). With SCA, 13 of 20 participants improve in batch retrieval and 19 of 20 in simulated streaming (\supp).

\subsection{Why the population works}
\label{sec:analyses}
We examine what the population shares with a new subject and how this helps calibration (BHA; details in the \supp).

\begin{figure}[!tb]
\centering
\includegraphics[height=.868\columnwidth]{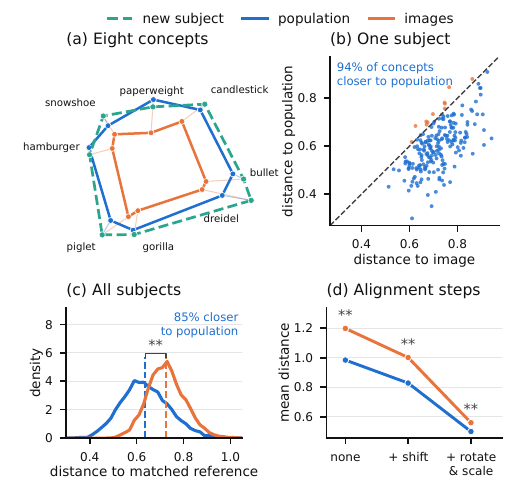}
\caption{\textbf{The population stays closer after alignment} (BHA + SCA; offline paired fits). (a) One subject and eight concepts, aligned by a shift, a rotation and a global scale, on a joint principal plane. (b) All 200 concepts after orthogonal alignment. (c) Distances to the matched reference; dashed: means. (d) Mean distance after each step.}
\label{fig:rigid}
\end{figure}

\paragraph{The population shares the new subject's deviation.}
The population agrees more closely with the new subject than images do, both in global geometry and in local neighborhoods (\cref{fig:geo}(a,b)).
What the population shares is a deviation from the images: the new subject's per-image EEG$-$image residual points the same way as the population's (cosine 0.58, as high as between two training subjects), and predicted-anchor residuals follow measured ones (0.80; \cref{fig:geo}(c)).
Measured after a shared encoder, this deviation may combine shared neural responses with the encoder's own stimulus-specific errors; the relay exploits it either way.
It also survives rigid alignment: after a shift and a rotation, the new subject is still closer to the population than to the images for 85\% of concepts, and the population remains closer at every alignment step (\cref{fig:rigid}(c,d)).

\paragraph{What remains is mostly a shift and a rotation.}
Fitting transforms from a new subject to the measured population on half of the test concepts and identifying the other half, a shift and a rotation raise identification from 53.7 to 89.4\%, close to a training subject against the other sources (97.7\%), whereas more flexible maps do worse.
Even with labels, aligning to the population generalizes to unseen concepts better than aligning to the images (66.4 vs.\ 60.0\%; \supp).

\paragraph{Measured population responses reveal what the predictor recovers.}
Initial unlabeled matching on BHA pairs 69\% of responses correctly with image embeddings, 74\% with predicted population anchors and 91\% with the measured source average for each image (\cref{fig:geo}(d)). The predicted anchors thus recover part of the shared response pattern that helps identify the stimulus. In a separate offline test, replacing predicted anchors with measured ones while holding the rest of calibration fixed raises final retrieval from 78.37 to 85.00\% (\cref{fig:relay}(b)). A deployable predictor shows the same effect. Adding patch-token residuals brings its anchors closer to the measured ones and improves retrieval on both backbones (\supp).

\paragraph{Source diversity improves the reference.}
At a fixed budget of 80 trials per image, measured anchors from eight people with ten repetitions each yield 83.19\% retrieval accuracy, versus 76.51\% from one person with eighty (\cref{fig:relay}(d)). Sampling more people better captures what recurs in a new subject. Predicted anchors show the same trend as the source pool grows from one to nine (71.69 to 78.37\%; \cref{fig:relay}(c)), consistent with \cref{eq:reference_error}.

\subsection{Where the gain comes from}
\label{sec:gain}

\begin{figure}[!tb]
\centering
\includegraphics[width=\columnwidth]{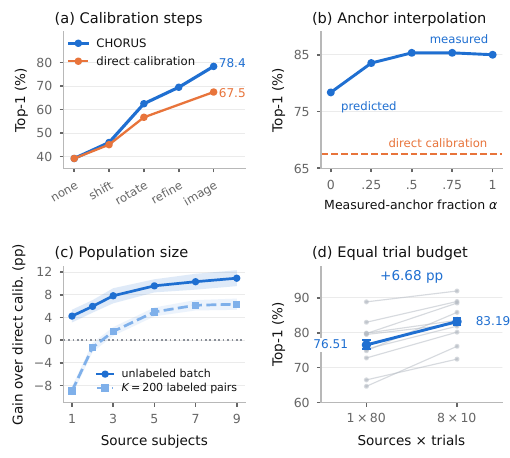}
\caption{\textbf{Population-relay analysis} (BHA). (a) Stage accuracy with a fixed CSLS readout. (b) Anchors interpolated from predicted ($\alpha=0$) to measured ($\alpha=1$). (c) Gain over direct by source count (seed SD). (d) Equal 80-trial source budget; gray: subjects, blue: mean $\pm$ seed SD. (b,d) use source test EEG offline.}
\label{fig:relay}\label{fig:lab}
\end{figure}

\paragraph{Population alignment gives image refinement a better starting point.}
On BHA, EEG--image pairing accuracy rises from 66.70\% to 78.23\% after population alignment. Starting from these more reliable pairs, image refinement raises retrieval accuracy from 69.48\% to 78.37\% under the same readout (\cref{fig:relay}(a)). The first-stage reference also matters for NSD fMRI. Changing only this reference from image embeddings to predicted population anchors improves 907-way retrieval for all four subjects, raising mean Top-1 from 70.51\% to 86.93\% (\cref{app:fmri}).

Across all four EEG decoders, most errors the relay corrects are near misses. The correct image moves into first place after narrowly trailing the top candidate under direct calibration, even though mean cosine similarity between queries and their correct images changes little (\supp).

\paragraph{The gain is not tied to one choice.}
The relay improves retrieval across all training schemes in \cref{tab:training}, with pooled-feature anchors, with cosine scoring, and on two additional decoders, Brain-HiVE and ATM-S. Removing population refinement causes the largest drop in the component ablation, reducing Top-1 by 6.60 points on BHA and 4.15 on SAMGA.

\begin{figure}[!tb]
\centering
\includegraphics[width=\columnwidth,trim=0 5 0 2,clip]{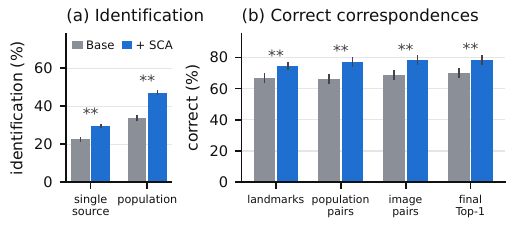}
\caption{\textbf{What SCA changes} (BHA, three seeds). (a) Identification of the held-out subject against one source or the population. (b) Correct transport landmarks, population- and image-stage pairs, and final Top-1. Bars: SEM; stars: paired tests over subjects; labels used only for evaluation.}
\label{fig:sca}
\end{figure}

\paragraph{SCA sharpens the correspondences the relay uses.}
With SCA, cross-subject identification against the measured population rises from 33.8 to 47.2\% (\cref{fig:sca}(a)). With predicted anchors, both stages infer more accurate pairs, and final Top-1 rises from 69.95 to 78.37\% (\cref{fig:sca}(b)).

\subsection{Robustness and limits}
\label{sec:robustness}
\paragraph{Fewer queries, a larger gallery, and fewer trials.}
With queries for only 50 or 100 of the 200 images, the relay still leads direct calibration by 4.6--8.7 points. The gain persists with 1{,}000 candidates and with only 4--20 trials per image (\supp).

\begin{table}[!ht]
\centering\footnotesize\setlength{\tabcolsep}{3pt}
\caption{\textbf{Retrieval with full and sparse matching.} Unlabeled 200-way Top-1 (\%), ten subjects and three seeds. CHORUS uses the same matcher in both stages.}
\label{tab:sparse_main}
\begin{tabular*}{\columnwidth}{@{\extracolsep{\fill}}lrrrr@{}}\toprule
& \multicolumn{2}{c}{Hungarian} & \multicolumn{2}{c}{Sparse MNN}\\
\cmidrule(lr){2-3}\cmidrule(l){4-5}
Decoder & Direct & CHORUS & Direct & CHORUS\\\midrule
BHA & 67.45 & \textbf{78.37} & 59.78 & \textbf{69.23}\\
SAMGA & 66.77 & \textbf{73.40} & 63.07 & \textbf{66.92}\\
\bottomrule\end{tabular*}
\end{table}

\paragraph{Sparse matching.}
With SCORE's mutual-nearest-neighbor (MNN) rule, CHORUS improves on direct calibration by 9.45 points on BHA and 3.85 on SAMGA (\cref{tab:sparse_main}). On SAMGA, MNN raises population-stage pair precision from 73.45 to 78.07\%, yet retains only 105 correct pairs per batch, down from 147. Randomly reducing the Hungarian pairs to the same count costs only 0.5 points (\supp), suggesting that which responses are retained also matters.

\begin{table}[!t]
\centering\small
\caption{\textbf{Calibration transfer to unseen images.} 100-way Top-1 (\%) for ten subjects and one seed, using five random splits in both directions. Frozen maps are reused from the other 100 images. Refit maps are estimated on the new batch.}
\label{tab:transfer_stages}
\begin{tabular*}{\columnwidth}{@{\extracolsep{\fill}}lllrr@{}}\toprule
Method & Population & Image & BHA & SAMGA\\
\midrule
CHORUS & Frozen & Frozen & 75.13 & 74.72\\
CHORUS & Refit & Frozen & 83.97 & 82.67\\
CHORUS & Frozen & Refit & 85.15 & 84.34\\
CHORUS & Refit & Refit & \textbf{85.98} & \textbf{85.16}\\
\midrule
Direct & --- & Frozen & 71.98 & 72.35\\
Direct & --- & Refit & 74.53 & 76.87\\
\bottomrule
\end{tabular*}
\end{table}

\paragraph{Transfer to unseen images.}
Does population alignment capture a shift that carries over when the same subject sees different images? We fit both stages without labels on 100 concepts and froze them for the other 100 (\cref{tab:transfer_stages}). The frozen relay outperforms frozen direct calibration for all ten subjects on both backbones. Reusing the population map and refitting only the image stage comes within one point of refitting both stages on each backbone. Thus the first stage gives image matching a useful starting point even on a new gallery, where the second stage can infer fresh correspondences.

\subsection{Image reconstruction}
\label{sec:reconstruction_results}

\begin{figure}[!ht]
\centering
\includegraphics[width=\columnwidth,trim=0 0 0 6,clip]{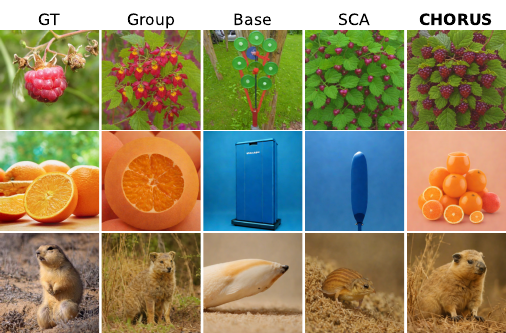}
\caption{\textbf{Held-out-subject reconstructions (BHA).} Each row shows one stimulus (GT) and matched reconstructions. Group decodes the measured population anchor for that test image (oracle). Target EEG and generation seeds are matched within rows.}
\label{fig:recon}
\end{figure}

We feed calibrated embeddings to a fixed reconstruction pipeline (a diffusion prior~\cite{ramesh2022hierarchical,scotti2023reconstructing} and SDXL-Turbo~\cite{sauer2024adversarial} with an IP-Adapter~\cite{ye2023ipadapter}).
To keep test images out of the conditioning, the deployable calibration here is a single stage toward targets built from training images: measured population anchors of the training concepts (population relay) or their image embeddings (direct calibration).
Population alignment slightly improves all six feature-based reconstruction metrics over direct calibration, including two-way CLIP identification (0.720 vs.\ 0.713). Direct calibration yields higher pixel correlation (0.182 vs.\ 0.174). The benefit beyond retrieval is clearest in the visual features recovered by the fixed generator (\cref{tab:reconstruction,fig:recon}).

\begin{table}[!ht]
\centering
\caption{\textbf{Cross-subject reconstruction (BHA).} Ten subjects, one seed. Metrics are pixel correlation, two-way identification (AlexNet, Inception, CLIP) and feature distance (EfficientNet, SwAV). Blue marks the oracle and bold the best deployable score.}
\label{tab:reconstruction}
\begingroup
\fontsize{8}{9.6}\selectfont
\setlength{\tabcolsep}{1pt}
\renewcommand{\arraystretch}{1.23}
\begin{tabular*}{\columnwidth}{@{\extracolsep{\fill}}p{62pt}ccccccc@{}}
\toprule
Configuration & {\scriptsize PixCorr$\uparrow$} & {\scriptsize Alex.2$\uparrow$} & {\scriptsize Alex.5$\uparrow$} & {\scriptsize Incep.$\uparrow$} & {\scriptsize CLIP$\uparrow$} & {\scriptsize Eff.$\downarrow$} & {\scriptsize SwAV$\downarrow$}\\
\midrule
\rowcolor{groupblue!8}
\textcolor{groupblue}{\textbf{Group}} & 0.281 & 0.895 & 0.943 & 0.836 & 0.880 & 0.809 & 0.506\\
\midrule
Base & 0.126 & 0.700 & 0.727 & 0.614 & 0.636 & 0.945 & 0.623\\
+ Diffusion prior & 0.122 & 0.689 & 0.716 & 0.610 & 0.644 & 0.945 & 0.637\\
+ SCA & 0.141 & 0.716 & 0.753 & 0.638 & 0.682 & 0.932 & 0.627\\
+ Direct calibration & \textbf{0.182} & 0.764 & 0.805 & 0.667 & 0.713 & 0.913 & 0.608\\
\midrule
\rowcolor{oursbg}
\textbf{CHORUS} & 0.174 & \textbf{0.769} & \textbf{0.815} & \textbf{0.684} & \textbf{0.720} & \textbf{0.911} & \textbf{0.606}\\
\bottomrule
\end{tabular*}
\endgroup
\par\vspace{4pt}
{\fontsize{7.8}{9.4}\selectfont\raggedright
\textcolor{groupblue}{Group}: source-average test embeddings (target excluded) decoded directly, with the target's spatial branch.
\par}

\end{table}

\subsection{Limitations}
\label{sec:experimental_scope}
CHORUS needs source subjects who viewed shared stimuli: both measured and predicted anchors are built from their responses to the same images, so datasets without shared stimuli do not provide a population reference directly.
\par\noindent
Our streaming is not yet closed-loop online use: it averages repeated trials under a known presentation schedule and replays recorded data.
We use public datasets with the original studies' consent and ethics approval; neural-data privacy remains important.

\section{Conclusion}
\label{sec:conclusion}
For a new subject, the population provides a closer reference than the images. CHORUS first corrects the subject-specific shift, then aligns the shared response to the images.
The decoder's visual features predict population anchors before the new subject arrives. SCA improves cross-subject matching by changing only the training data.
The relay beats direct calibration at every budget and on both datasets, and on all four decoders under sparse matching; SCA and direct calibration also beat prior schemes on their own.
Better image-to-population prediction, open-ended galleries and live sessions come next.

\section*{LLM Usage Statement}
We used Claude through Claude Code and OpenAI Codex to assist with research development, implementation, and preparation of the main paper and supplement. Their use included discussing experimental designs and interpretations; writing and debugging training, evaluation, analysis, and plotting code; launching and monitoring runs; aggregating and checking recorded results; and drafting, editing, and critiquing text, captions, and related-work discussions. The authors remain responsible for the methods, evidence, claims, citations, and final submission. We used SDXL-Turbo with IP-Adapter to generate the reconstructions (\cref{sec:reconstruction_results}).

{\small
\bibliographystyle{ieeenat_fullname}
\bibliography{references}
}
\clearpage
\appendix
\setcounter{figure}{0}
\setcounter{table}{0}
\setcounter{equation}{0}
\renewcommand{\thefigure}{S\arabic{figure}}
\renewcommand{\thetable}{S\arabic{table}}
\renewcommand{\theequation}{S\arabic{equation}}
\renewcommand{\theHfigure}{supp.\arabic{figure}}
\renewcommand{\theHtable}{supp.\arabic{table}}
\renewcommand{\theHequation}{supp.\arabic{equation}}
\raggedbottom
\renewcommand{\topfraction}{.92}
\renewcommand{\bottomfraction}{.85}
\renewcommand{\textfraction}{.07}
\renewcommand{\floatpagefraction}{.8}
\renewcommand{\dbltopfraction}{.92}
\renewcommand{\dblfloatpagefraction}{.8}
\setcounter{topnumber}{3}
\setcounter{bottomnumber}{2}
\setcounter{totalnumber}{5}
\setcounter{dbltopnumber}{3}
\setlength{\textfloatsep}{9pt plus 2pt minus 2pt}
\setlength{\dbltextfloatsep}{9pt plus 2pt minus 2pt}

\newenvironment{overviewtable}{\par\medskip\begin{minipage}{\linewidth}\captionsetup{type=table}}{\end{minipage}\par}
\twocolumn[{%
\begin{minipage}{\textwidth}
\centering{\Large\bfseries Supplementary Material\par}\vspace{1em}
\raggedright
\section{Overview}
\label{app:overview}
This supplement follows the order of the main paper: method details (\cref{app:method}), experimental setup (\cref{app:setup}), additional retrieval results (\cref{app:results}), additional analyses (\cref{app:analysis}), the calibrator across four decoders (\cref{app:anatomy}), image reconstruction (\cref{app:reconstruction}) and an extension to fMRI (\cref{app:fmri}). \Cref{tab:s_map} lists where each pointer in the main text is answered, and \cref{tab:s_access} states the information every experiment uses. Code, checkpoints, per-subject results and the selected configurations will be released.

\begin{overviewtable}
\centering\footnotesize\setlength{\tabcolsep}{5pt}
\caption{Where each pointer in the main text is answered.}
\label{tab:s_map}
\begin{tabular}{@{}p{.25\textwidth}p{.55\textwidth}l@{}}\toprule
Main text & Topic & Here\\\midrule
\S3.3 & Anchor predictor: inputs, patch-token residual, source-validation choices & \cref{app:anchor}\\
\S3.3, Eq.~(2) & Derivation of the reference-error model & \cref{app:refmodel}\\
\S3.4, Eq.~(4) & Regularized orthogonal recovery and its polar-factor solution & \cref{app:recovery}\\
\S3.4--3.5 & Unlabeled correspondence inference, labeled support, causal streaming & \cref{app:correspondence,app:labeled,app:s_stream_steps}\\
\S4.1, Tabs.~1--2 & Data, statistics, decoders, SCA recipe, SATTC/SCORE interfaces & \cref{app:setup}\\
\S4.2, Fig.~4 & Labeled support at every budget & \cref{app:kshot_results}\\
\S4.2, \S4.4, Tab.~1 & Training schemes under each route, pooled-feature anchors, component deletions, cosine readout & \cref{app:components}\\
\S4.1--4.2, Tabs.~1--2 & Baseline training protocol; where the prior schemes lose accuracy & \cref{app:baselines,app:rivals}\\
\S4.5 & Fewer queries, larger gallery, fewer trials (incl.\ SAMGA) & \cref{app:robust}\\
\S4.5, Tab.~4 & Sparse matching (MNN) and the pair-count control & \cref{app:s_matchers}\\
\S4.5, Tab.~5 & Frozen-stage transfer and unlabeled calibration budget & \cref{app:half}, \cref{fig:s_holdout_k}\\
\S4.2, Tab.~3 & Alljoined-1.6M batch and simulated stream & \cref{app:xdataset}\\
\S1, \S4.3, Figs.~1(a,b), 5--6 & Geometry, per-image cosine, subject-level correlation & \cref{app:geometry}\\
\S4.3--4.4, Fig.~7 & Calibration stages, anchor interpolation, source count, equal trials & \cref{app:stages,app:interp,app:s_diversity}\\
\S4.3, \S4.5 & Transform family, rotation rank, held-out retrieval, frozen calibration on new images & \cref{app:half}\\
\S4.4, Fig.~8 & What SCA changes; fixed-update control; layer-wise analysis & \cref{app:s_layerwise}\\
\S4.3--4.4 & The calibrator on four decoders: stages, encoder geometry, correspondences, labeled rotation, where queries land & \cref{app:anatomy}\\
\S4.6, Tab.~6, Fig.~9 & Reconstruction protocol, all metrics, examples & \cref{app:reconstruction}\\
\S3.3, \S4.3--4.4 & Population geometry and two-stage calibration in fMRI & \cref{app:fmri}\\\bottomrule
\end{tabular}
\end{overviewtable}

\begin{overviewtable}
\centering\footnotesize\setlength{\tabcolsep}{4pt}
\caption{Information used by each experiment. Population references always exclude the target subject.}
\label{tab:s_access}
\begin{tabular}{@{}p{.17\textwidth}p{.28\textwidth}p{.28\textwidth}p{.21\textwidth}@{}}\toprule
Setting & Target observations used to fit calibration & Reference & Readout\\\midrule
Labeled support & $K$ training-image EEG--anchor pairs & Measured source-training anchors (fit); predicted test anchors (gallery) & Support-CSLS; each test query scored on its own\\
Unlabeled batch & 200 test responses jointly, without labels & Predicted anchors of the candidate images; candidate image embeddings & Batch-CSLS on the same 200 queries\\
Causal stream & Earlier arrivals with repetition-group IDs & Predicted anchors and image embeddings & History-CSLS; prediction before history update\\
Reconstruction & 200 unlabeled test responses & Measured anchors or image embeddings of training concepts & Fixed generator; no test images as conditions\\
Offline diagnostics & As stated per analysis & May use source test EEG and labels of fitting concepts & Reported apart from deployable results\\\bottomrule
\end{tabular}
\end{overviewtable}

\end{minipage}\vspace{\textfloatsep}
}]
\section{Method details}
\label{app:method}

\subsection{Anchor predictor}
\label{app:anchor}\label{app:patch_residual}
\begin{figure*}[t]\centering
\includegraphics[width=\textwidth]{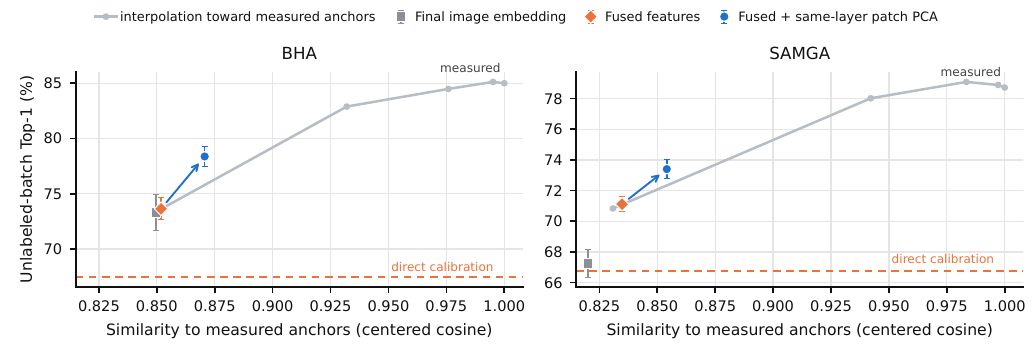}
\caption{\textbf{Anchor similarity and retrieval for three predictors} (unlabeled batch, three seeds). Points: three-seed mean; bars: SD of the seed means. Gray: offline interpolation from predicted toward measured anchors (main Fig.~7b); orange dashed: direct calibration. Measured anchors use source test EEG as an offline reference.}
\label{fig:s_anchorsim}\end{figure*}
\begin{table*}[t]
\centering\footnotesize\setlength{\tabcolsep}{4pt}
\caption{Anchor-predictor inputs with fixed encoders and calibration. Ridge penalties use source-validation cosine; test-anchor metrics use source test EEG offline. Top-1: three-seed mean $\pm$ SD (\%); $R^2$: centered measured anchors. Bold: best per backbone and metric.}
\label{tab:s_anchorq}
\begin{tabular*}{\textwidth}{@{\extracolsep{\fill}}llrrr@{}}\toprule
Backbone & Predictor input & Anchor cosine & $R^2$ & Batch Top-1\\\midrule
BHA & Final image embedding & 0.850 & 0.709 & $73.28\pm1.64$\\
BHA & Original visual features & 0.835 & 0.675 & $74.27\pm1.12$\\
BHA & Joint (fused features) & 0.852 & 0.706 & $73.63\pm0.98$\\
BHA & Joint + patch-token residual (CHORUS) & \textbf{0.871} & \textbf{0.743} & $\mathbf{78.37}\pm0.90$\\
SAMGA & Final image embedding & 0.820 & 0.570 & $67.25\pm0.89$\\
SAMGA & Original visual features & 0.831 & 0.572 & $70.83\pm0.90$\\
SAMGA & Joint (fused features) & 0.835 & 0.586 & $71.12\pm0.47$\\
SAMGA & Joint + patch-token residual (CHORUS) & \textbf{0.854} & \textbf{0.627} & $\mathbf{73.40}\pm0.61$\\
Brain-HiVE & Final image embedding & 0.820 & 0.668 & $47.35\pm1.05$\\
Brain-HiVE & Original visual features & 0.787 & 0.606 & $49.00\pm1.79$\\
Brain-HiVE & Joint (fused features) & 0.836 & 0.650 & $49.92\pm1.39$\\
Brain-HiVE & Joint + patch-token residual (CHORUS) & \textbf{0.858} & \textbf{0.713} & $\mathbf{54.03}\pm1.86$\\
\bottomrule\end{tabular*}
\end{table*}

A measured anchor averages the frozen encoder's source embeddings of an image. For candidate images that the sources never saw, a predictor maps image features to anchors in two steps. It reads only the vision encoders and layers that the decoder itself uses, is fitted on source-training images, and makes every choice by the centered cosine to measured anchors on source-validation images. The 200 test anchors are fixed before any target data are used.

\paragraph{Pooled features.}
Let $M$ stack the decoder's pooled visual features of the source-training images, each row-normalized before concatenation, and $V$ their final image embeddings. Each block is centered and divided by its root-mean-square row norm on the training images, $\widetilde M=(M-\bar M)/\sqrt{\|M-\bar M\|_F^2/N}$, and likewise $\widetilde V$. With centered measured anchors $A_c$, a ridge regression with separate penalties solves
\begin{align}
\min_{W_m,W_v}\;&\|A_c-\widetilde M W_m-\widetilde V W_v\|_F^2\notag\\
&+\lambda_m\|W_m\|_F^2+\lambda_v\|W_v\|_F^2,
\label{eq:s_joint}
\end{align}
and predicts $\widehat a_{\rm joint}=\bar a+\widetilde mW_m+\widetilde vW_v$, with $\lambda_m\in\{.001,.01,.1,1,10,20\}$ and $\lambda_v/\lambda_m\in\{.01,.1,1,10,100\}$. The pooled block holds BHA's four visual encoders, SAMGA's five InternViT layers, and Brain-HiVE's SynCLR, CLIP and VAE features. ATM-S has a single OpenCLIP feature and uses one ridge map.

\paragraph{Patch-token residual.}
Pooling discards where things lie in the image. A second ridge regression therefore predicts the residual $A-\widehat A_{\rm joint}$ from the patch tokens of the same layers: the final OpenCLIP ViT-H/14 layer, DINOv2-L layer 12, MAE layer 18 and DFN layer 6 for BHA; InternViT-6B layers 20, 24, 28, 32 and 36 for SAMGA; the final SynCLR layer for Brain-HiVE; and the final OpenCLIP layer for ATM-S. Tokens are taken after the layer's final normalization, $\ell_2$-normalized, average-pooled to at most $16\times16$ (DFN keeps its $7\times7$ grid) and reduced by PCA fitted on source-training images to 256 or 1,024 components per layer; the components of all layers are concatenated and scaled to unit mean row norm. Training residuals use cross-fitted pooled predictions, so the second regression does not learn the first one's in-sample fit. Source validation chooses, per fold, the number of components, linear or degree-4 polynomial kernel PCA, whether to standardize components, the penalty $\lambda\in\{0.01,\dots,1000\}$, and residual or direct prediction; a fold keeps the pooled predictor if no variant improves the validation cosine. The prediction is $\widehat a=\widehat a_{\rm joint}+\widehat r$.

\paragraph{Effect.}
\Cref{tab:s_anchorq} compares predictor inputs with encoders and calibration fixed. Among pooled predictors, the joint ridge has the highest anchor cosine on all three multi-feature decoders, although a higher cosine does not always give higher accuracy (BHA: 74.27\% with the original visual features alone, 73.63\% jointly). The patch-token residual is best on all three measures for all three decoders: the centered cosine rises from 0.85 to 0.87 on BHA and from 0.83 to 0.85 on SAMGA, and batch accuracy from 73.63 to 78.37\% and from 71.12 to 73.40\%, for every held-out subject. \Cref{fig:s_anchorsim} places the predictors next to the offline interpolation toward measured anchors: at equal similarity to the measured anchors, the patch-token predictor retrieves better than interpolated anchors.

\subsection{Regularized orthogonal recovery}
\label{app:recovery}
For matched rows $(x_i,y_i)$ with weights $w_i\ge0$, $\sum_iw_i=1$, let $X_c,Y_c$ be centered at their weighted means and $F=X_c^\top\operatorname{diag}(w)Y_c$. Recovery solves
\begin{equation}
\min_{R^\top R=I}\sum_iw_i\|x_{c,i}R-y_{c,i}\|^2+\rho\sigma\|R-I\|_F^2,
\label{eq:s_polar}
\end{equation}
with $\sigma=\|F\|_2$. Expanding the squares leaves the maximization of $\operatorname{tr}[R^\top(F+\rho\sigma I)]$, solved by the polar factor $R=UV^\top$ of $F+\rho\sigma I=U\Sigma V^\top$~\cite{schonemann1966generalized}. If $B$ has orthonormal columns spanning the centered pairs, then $F=BB^\top FBB^\top$ and
\begin{equation}
R=B\,\operatorname{polar}(B^\top FB+\rho\sigma I_r)B^\top+I-BB^\top ,
\label{eq:s_span}
\end{equation}
so directions outside the span of the pairs stay unchanged and only an $r\times r$ problem is solved. $T(h)=(h-\mu_x)R+\mu_y$ preserves pairwise distances; larger $\rho$ pulls $R$ toward the identity, which stabilizes maps fitted on few pairs. Transport initialization weights pairs by their coupling mass; subsequent discrete matches use uniform weights. When correspondences are re-inferred after a provisional transform, recovery is refitted from the original shifted target embeddings, so transforms do not accumulate across iterations.

\subsection{Unlabeled correspondence inference}
\label{app:correspondence}
\begin{algorithm}[t]
\small
\textbf{Input:} batch $H$, fixed predicted anchors $\widehat A$, gallery $V$, source-population mean $\mu_a$.
\begin{enumerate}[leftmargin=*,itemsep=1pt,topsep=2pt]
\item Shift $Q\leftarrow H-\bar H+\mu_a$.
\item Initialize $T_1$ from transport landmarks between $Q$ and $\widehat A$ (Eq.~\ref{eq:s_polar}).
\item Alternate assignment from $T_1(Q)$ to $\widehat A$ and refitting $T_1$ from $Q$.
\item Freeze $Z=T_1(Q)$. Alternate assignment from $T_2(Z)$ to $V$ and moment-matched recovery from $Z$.
\item Score $T_2(Z)$ against $V$ once with batch CSLS; return each row's maximum.
\end{enumerate}
\caption{Unlabeled population-relay calibration.}
\label{alg:s_batch}
\end{algorithm}
The target batch is first shifted to the source-population mean. Correspondences are initialized by an FGW-inspired transport iteration~\cite{vayer2019optimal}. With $D_{qa}=1-\cos(Q,\widehat A)$ and within-set cosine distances $D_q,D_a$, each normalized by its mean, the linearized cost at coupling $\Pi$ is
\begin{align}
L(\Pi)&=(1-\alpha)D_{qa}\notag\\
&+\alpha[c_q\mathbf1^\top+\mathbf1c_a^\top-2D_q\Pi D_a^\top],
\end{align}
where $c_q=D_q^{\odot2}p_q$ and $c_a=D_a^{\odot2}p_a$. We use log-domain entropic scaling with an exact query marginal and a KL-relaxed candidate marginal ($\alpha=.5$, column penalty $\tau=1$, 30 outer and 60 inner iterations, entropy annealed from $10\epsilon$ to $\epsilon=.05$ over the first two thirds). Mutual coupling maxima give up to 160 landmarks; recovery starts once at least 12 exist. We then alternate maximum-cosine Hungarian assignment~\cite{kuhn1955hungarian} with unweighted recovery ($\rho=.1$) until the assignment repeats or 100 iterations are reached. Each recovery is refitted from the original shifted input, not composed with the previous one; rectangular assignment is used when there are fewer queries than candidates.

The image stage starts from the fixed population-aligned representation, infers a fresh assignment to the image embeddings, matches per-coordinate means and standard deviations of the inferred pairs, and fits regularized recovery with $\rho=.1$, again refitting from its fixed input (\cref{alg:s_batch}). Direct calibration shifts toward the gallery mean, initializes transport against the image embeddings and applies only this image stage.

\paragraph{CSLS.}
For a cosine matrix $S$, CSLS~\cite{conneau2018word} returns $2S_{ij}-r_i-c_j$, where $r_i$ averages the ten largest scores of query $i$ and $c_j$ the strongest query scores of candidate $j$: from the test batch, from earlier groups in a stream, or from the mapped supports with labels. Only $c_j$ can change a query's ranking. ATM-S keeps its unnormalized gallery vectors in both routes.

\subsection{Labeled support}
\label{app:labeled}
Let the $K$ supports be $(x_i,a_i)$, target embeddings and measured anchors of the same training images. With support means $\bar x,\bar a$ and per-coordinate variances $v_x,v_a$, a single global scale
\begin{equation}
s=\frac{\sqrt{\operatorname{mean}(v_a)}+10^{-6}}{\sqrt{\operatorname{mean}(v_x)}+10^{-6}}
\end{equation}
matches the overall spread, and $R_u$ solves Eq.~\ref{eq:s_polar} on the scaled, centered pairs with uniform weights and $\rho=10/K$, so that
\begin{equation}
T_{{\rm pop},K}(h)=s\,(h-\bar x)R_u+\bar a .
\end{equation}
Each test query is then scored on its own against the 200 predicted anchors of the candidate images; the hubness term of each candidate uses its $\min(10,\max(2,\lfloor K/2\rfloor))$ strongest scores from the mapped supports, never other test queries. Direct calibration fits the identical map to the image embeddings of the same supports and scores the image gallery with the same readout. $\rho=10/K$ and the global scale are fixed in advance for every fold, seed, draw and budget.

\begin{algorithm}[t]
\small
\textbf{Before target evaluation:} fit the anchor predictor on source data; predict the 200 test anchors.
\begin{enumerate}[leftmargin=*,itemsep=1pt,topsep=2pt]
\item Fit $s$, $R_u$ on the $K$ supports with $\rho=10/K$.
\item Map supports and queries through $T_{{\rm pop},K}$.
\item Compute each candidate's hubness term from the mapped supports.
\item Return, per query, the candidate whose predicted anchor has the highest support-CSLS score.
\end{enumerate}
\caption{CHORUS with labeled support. Direct calibration fits steps 1--2 toward image embeddings and scores the image gallery.}
\label{alg:s_kshot}
\end{algorithm}

\subsection{Causal streaming}
\label{app:s_stream_steps}
On THINGS-EEG2 each subject yields 16,000 arrivals (200 images, 80 repetitions) in four blocks of 20 repetitions, randomized within blocks; three arrival orders are shared by all methods. A repetition-group ID links trials of the same, unidentified image, so the current trial is averaged with earlier trials of its group before encoding. The history keeps one embedding $h_g$ per observed group. At each refit, every 200 arrivals once at least 12 groups exist, the query mean is shrunk toward the source mean,
\begin{equation}
\mu_t=\frac{8\mu_a+\sum_{g}h_g}{8+n},
\end{equation}
and both stages of \cref{alg:s_batch} are refitted on $Q=H-\mu_t+\mu_a$ with rectangular assignments. The gallery correction uses cached scores of other groups only, and a group's history entry is replaced only after its prediction (\cref{alg:s_stream}). Direct calibration in the stream follows the same schedule, prior, and scoring, but centers the history on the gallery mean ($Q=H-\mu_t+\bar V$) and aligns it to the images in one step before the same gallery correction.

\begin{algorithm}[t]\small
\textbf{Input:} arrivals with group IDs; encoder $f$; fixed anchors $\widehat A$; gallery $V$; source mean $\mu_a$.
\begin{enumerate}[leftmargin=*,itemsep=1pt,topsep=2pt]
\item Start with an empty history and the identity transform.
\item Before arrivals 201, 401, \ldots, compute $\mu_t$; with at least 12 groups, refit both stages and refresh cached scores.
\item Encode the running mean waveform of the current group, including the arriving trial.
\item Transform, score against $V$ with the CSLS correction from other cached groups, and return the best image.
\item Then replace the group's history entry and cached score.
\end{enumerate}
\caption{Causal streaming. Fitting and the gallery correction use only earlier arrivals.}
\label{alg:s_stream}
\end{algorithm}

\subsection{Constants and cost}
\label{app:constants}
\Cref{tab:s_constants} lists the settings shared across subjects.
\begin{table}[t]\centering\footnotesize
\setlength{\tabcolsep}{3pt}
\caption{Calibration constants. None is selected on target data.}
\label{tab:s_constants}
\begin{tabular}{@{}p{.39\columnwidth}p{.55\columnwidth}@{}}\toprule
Component & Setting\\\midrule
Pooled anchor ridge & 30 source-validation candidates; Eq.~\ref{eq:s_joint}\\
Patch-token residual & Same layers; PCA 256/1,024; source-validation choices\\
Transport & $\alpha=.5$, $\tau=1$, $\epsilon=.05$; 30 outer, 60 inner iterations\\
Transport landmarks & At least 12; at most 160\\
Unlabeled recovery & $\rho=.1$ in both stages\\
Assignment refinement & Stop on repetition; at most 100 per stage\\
Batch / stream CSLS & Up to 10 neighbors\\
Streaming & Prior weight 8; 12-group warm-up; refit every 200 arrivals\\
Labeled population stage & Global scale, shift, rotation; $\rho=10/K$\\
Labeled readout & Predicted anchors (CHORUS); images (direct)\\
Support CSLS & $\min(10,\max(2,\lfloor K/2\rfloor))$ neighbors\\\bottomrule
\end{tabular}
\end{table}
For $n$ queries, $m$ candidates and dimension $d$, the score matrix costs $O(nmd)$ and square Hungarian assignment $O(n^3)$. With $l$ fitting pairs, Eq.~\ref{eq:s_span} needs a basis of size $r\le\min(d,2l)$, costing $O(dlr)$ to form and $O(r^3)$ for the polar factor, so no dense $d\times d$ decomposition is needed when $l\ll d$. Labeled queries reuse the fitted map; streams recalibrate only at scheduled refits.

\subsection{Reference-error model}
\label{app:refmodel}
After rigid alignment, write $h_{s,i}=v_i+\delta_i+e_{s,i}$ with a shared deviation $\delta_i$ and independent zero-mean subject terms of power $\eta^2$. For image $i$, a new subject $u$ and $S$ independent sources,
\begin{align}
\mathbb E\|h_{u,i}-v_i\|^2&=\|\delta_i\|^2+\eta^2,\\
\mathbb E\|h_{u,i}-h_{s,i}\|^2&=2\eta^2,\\
\mathbb E\|h_{u,i}-\bar h_{\mathcal S,i}\|^2&=\eta^2+\eta^2/S,
\end{align}
since independence and zero means remove the cross terms. The population is closer than a single source whenever $S>1$, and closer than the image when $\|\delta_i\|^2>\eta^2/S$. With correlated source errors the population term becomes $S^{-2}\sum_{p,q}\mathbb E[e_{p,i}^\top e_{q,i}]$, so the model predicts orderings rather than the size of the gain.

\section{Experimental setup}
\label{app:setup}\label{app:implementation}

\subsection{Data, folds and statistics}
We use the released preprocessed THINGS-EEG2 recordings~\cite{gifford2022large}: 63 channels at 250\,Hz covering 0--1,000\,ms. The 16,540 training images comprise 1,654 concepts with ten exemplars and four repetitions each; the 200 test images belong to disjoint concepts and have 80 repetitions. A fixed concept-disjoint split reserves 165 training concepts for validation (14,890/1,650 images); encoder training, anchor regression, calibration choices and the reconstruction heads all use it. In each leave-one-subject-out fold, the target is absent from training and selection and the other nine subjects form the source population. Support responses average four trials and test queries 80.

Labeled supports use five disjoint draws per subject and budget, nested across budgets and shared by all methods and training seeds. Reported means first average draws or arrival orders and the ten subjects within each of three training seeds, then the three seed means; $\pm$ values are the SD of the seed means. Paired tests are two-sided Wilcoxon signed-rank tests over the ten subject means, after averaging seeds and draws, with Holm correction within each family (ten backbone--budget comparisons for labeled support; four decoders per matcher). Alljoined-1.6M and reconstruction use one training seed.

\subsection{Decoders and SCA}
\label{app:sca}
\paragraph{SCA.}
SCA averages the waveforms of source subjects who viewed the same image before encoding. Every epoch contains each individual row, one pair per individual row, a second pair for a random half of the rows and a quadruple for a random quarter (individual:pair:quadruple $=4:6:1$, i.e.\ 2.75 times the individual rows). Donors are resampled from the other sources at each access, and the anchor row supplies the subject identity that the backbone needs. Labels, losses and optimizers are unchanged. SCA averages waveforms for training, whereas population anchors average encoded responses at deployment.

\paragraph{BHA.}
We use the first training stage of BHA~\cite{anon2027bha}: a BrainViT EEG encoder with electrode tokens, electrode identities and montage positions, producing 1,024-dimensional embeddings, and a visual branch that fuses frozen OpenCLIP ViT-H/14, MAE, DINOv2-L and DFN features. Training uses the symmetric contrastive loss, AdamW (learning rate $3\times10^{-4}$), batch size 1,024 and validation-loss early stopping (patience ten, at most 100 epochs). The subject token is disabled for leave-one-subject-out training.

\paragraph{SAMGA.}
We use SAMGA's TSConv EEG encoder and five InternViT layers~\cite{samga2026}, with the subject-aware router in training and the global route at test time. Twenty epochs of coarse alignment with contrastive and decaying MMD terms precede fine alignment; AdamW uses learning rate and weight decay $10^{-4}$, batch size 1,024 and temporal smoothing, with source-validation Top-1 early stopping (patience ten after the coarse stage, at most 100 epochs).

\paragraph{Brain-HiVE and ATM-S.}
Brain-HiVE~\cite{zheng2026learning} uses the authors' shallow MLP on 17 occipito-parietal channels and a fused SynCLR-B/16, CLIP-B/32 and SDXL-VAE token; its SCA arm replaces one quarter of the rows with pairs, and source validation selects learning rates of $10^{-4}$ (SCA) and $2.5\times10^{-4}$ (Base). ATM-S~\cite{li2024visual} uses its official retrieval architecture with OpenCLIP ViT-H/14 and the standard SCA recipe. Both use the shared split, source-only checkpoint selection and three seeds.

\subsection{Baselines}
\label{app:baselines}
SATTC~\cite{huang2026sattc} and SCORE~\cite{cui2026score} are evaluated as complete systems with the authors' training and calibration code: SATTC with its own trained encoder, whose image targets are candidate-whitened, and SCORE with its own training objective (multi-positive alignment and the recovery term) on both backbones. SATTC's full training recipe (cosine logits with a learned temperature, subject dropout and its learning rate) lowered accuracy on both backbones in our runs, so its encoder keeps each backbone's recipe with the whitened targets. To compare calibration alone, we also apply their batch and streaming calibration to the SCA encoder; in streams they receive the same history as CHORUS. SATTC's candidate whitening is applied only where the encoder was trained on whitened targets; the SCA encoder was not, so SATTC calibrates it in the original gallery space. Neither method uses paired labels, so labeled support is reported only for the population relay and for direct calibration, which fits the known pairs with the same map.

Every baseline encoder uses the same data split, source subjects, three training seeds and source-validation checkpoint selection as ours, and keeps its backbone's optimizer, schedule and early-stopping rule; no baseline setting is chosen on target data, and complete systems always use their own training and calibration, never a stronger encoder. As a check of our SCORE implementation, SAMGA trained with it reproduces the accuracy SCORE's authors report for this backbone within one point, before and after calibration. \Cref{app:rivals} analyzes where the prior schemes lose accuracy.

To compare training alone, each backbone is also trained with SATTC's, SCORE's and ProCA's~\cite{zhou2026proca} objectives on the individual (Base) rows, with each backbone's own optimizer, schedule and source-validation selection, and the resulting encoders are evaluated without calibration (main Table~1) and with the same direct and population-relay routes as SCA, with anchor predictors fitted on source training images and chosen on source validation images (\cref{app:components}). ProCA has no released code and no test-time component; we implement its confusion-guided contrastive loss and structure-consistent interpolation from the paper (\(\beta=0.99\), \(\alpha\) rising from epoch 25 to \(\alpha_{\max}=0.8\) at epoch 40, weight 0.3 after a 10-epoch warm-up, Beta(0.4, 0.4) interpolation). Details the paper leaves open were fixed before any result: confusion reweights each negative logit by its row-centered class confusion at the backbone's initial temperature; class prototypes are the backbone's own image targets; interpolation acts on the per-electrode tokens before the BrainViT readout and on the temporal-block output of TSConv.

\section{Additional retrieval results}
\label{app:results}\label{app:batch_results}

\subsection{Labeled support at every budget}
\label{app:kshot_results}
\begin{table*}[t]
\centering\footnotesize\setlength{\tabcolsep}{4pt}
\caption{Labeled-support Top-1 (\%), mean $\pm$ SD of three seed means, each averaging ten subjects and five draws. All routes fit the same scale, shift, and rotation on the same supports. Direct targets and scores images; CHORUS targets measured anchors and scores predicted anchors, from the fused-feature ridge alone or with the patch-token residual (CHORUS). $\Delta$, wins, and Holm-adjusted paired tests (ten comparisons) compare CHORUS with direct calibration on subject means. Bold: best per row.}
\label{tab:s_kshot}
\begin{tabular*}{\textwidth}{@{\extracolsep{\fill}}lrrrrrrr@{}}\toprule
Backbone & $K$ & Direct & CHORUS, fused anchors & CHORUS & $\Delta$ (pp) & Wins & $p_{\mathrm{Holm}}$\\\midrule
BHA & 10 & $36.83\pm0.53$ & $37.98\pm0.76$ & $\mathbf{41.57}\pm0.70$ & +4.74 & 10/10 & 0.0195\\
BHA & 20 & $39.84\pm0.58$ & $41.43\pm0.57$ & $\mathbf{45.23}\pm0.68$ & +5.39 & 10/10 & 0.0195\\
BHA & 50 & $45.26\pm0.77$ & $47.14\pm0.65$ & $\mathbf{51.36}\pm0.65$ & +6.10 & 10/10 & 0.0195\\
BHA & 100 & $49.13\pm0.79$ & $51.52\pm0.60$ & $\mathbf{56.06}\pm0.92$ & +6.93 & 10/10 & 0.0195\\
BHA & 200 & $54.08\pm0.99$ & $55.78\pm0.39$ & $\mathbf{60.41}\pm0.73$ & +6.33 & 10/10 & 0.0195\\
SAMGA & 10 & $37.14\pm0.10$ & $38.74\pm0.45$ & $\mathbf{41.34}\pm0.65$ & +4.20 & 10/10 & 0.0195\\
SAMGA & 20 & $39.96\pm0.18$ & $41.97\pm0.30$ & $\mathbf{44.44}\pm0.60$ & +4.48 & 9/10 & 0.0195\\
SAMGA & 50 & $44.79\pm0.50$ & $46.83\pm0.42$ & $\mathbf{49.54}\pm0.71$ & +4.75 & 10/10 & 0.0195\\
SAMGA & 100 & $48.92\pm0.26$ & $50.54\pm0.41$ & $\mathbf{53.48}\pm0.33$ & +4.56 & 9/10 & 0.0195\\
SAMGA & 200 & $52.81\pm0.16$ & $54.35\pm0.28$ & $\mathbf{57.43}\pm0.24$ & +4.62 & 9/10 & 0.0195\\
\bottomrule\end{tabular*}
\end{table*}

\Cref{tab:s_kshot} reports all five budgets. CHORUS leads direct calibration by 4.74 to 6.93 points on BHA and 4.20 to 4.75 on SAMGA; nine or ten of ten subjects improve in every cell, and CHORUS is higher in every seed. The middle column isolates the anchor predictor: with pooled-feature anchors the same route gains 1.15 to 2.39 points on BHA and 1.54 to 2.04 on SAMGA, and the patch-token residual adds 2.5 to 4.6 points in every cell.

\subsection{Four decoders and every subject}
\label{app:decoders}
\begin{table*}[t]
\centering\footnotesize\setlength{\tabcolsep}{4pt}
\caption{Unlabeled-batch Top-1 (\%), three-seed mean $\pm$ SD. H: Hungarian; M: mutual nearest neighbors on CSLS-corrected similarities, as in SCORE. Paired letters denote population/image matchers. All routes apply CSLS to their backbone-specific similarities. All four decoders use the patch-token anchor predictor with their own vision layers. Bold: higher mean within each matcher pair.}
\label{tab:s_backbones}
\begin{tabular*}{\textwidth}{@{\extracolsep{\fill}}lrrrr@{}}\toprule
Backbone & Direct (H) & CHORUS (HH) & Direct (M) & CHORUS (MM)\\\midrule
BHA & $67.45\pm1.19$ & $\mathbf{78.37}\pm0.90$ & $59.78\pm1.08$ & $\mathbf{69.23}\pm1.78$\\
SAMGA & $66.77\pm0.25$ & $\mathbf{73.40}\pm0.61$ & $63.07\pm0.51$ & $\mathbf{66.92}\pm1.11$\\
Brain-HiVE & $45.75\pm0.97$ & $\mathbf{54.03}\pm1.86$ & $43.05\pm0.97$ & $\mathbf{50.07}\pm0.77$\\
ATM-S & $9.63\pm0.32$ & $\mathbf{16.32}\pm1.21$ & $8.43\pm0.64$ & $\mathbf{14.48}\pm1.33$\\
\bottomrule\end{tabular*}
\end{table*}

\begin{table*}[t]
\centering\footnotesize\setlength{\tabcolsep}{4pt}
\caption{Unlabeled-batch accuracy by held-out subject, averaged over three training seeds. D: direct; C: CHORUS. Every paired subject mean favors CHORUS on each of the four decoders.}
\label{tab:s_subjects}
\begin{tabular*}{\textwidth}{@{\extracolsep{\fill}}lrrrrrrrr@{}}\toprule
Subject & BHA D & BHA C & SAMGA D & SAMGA C & HiVE D & HiVE C & ATM D & ATM C\\\midrule
S1 & 78.67 & 88.00 & 75.67 & 81.17 & 55.17 & 62.67 & 6.83 & 14.00\\
S2 & 74.83 & 82.83 & 71.17 & 76.83 & 53.17 & 58.83 & 10.33 & 16.83\\
S3 & 70.33 & 87.17 & 63.50 & 72.00 & 40.33 & 52.33 & 10.33 & 19.00\\
S4 & 54.17 & 63.17 & 54.50 & 60.67 & 36.83 & 44.00 & 8.17 & 12.33\\
S5 & 62.67 & 72.83 & 67.33 & 74.50 & 45.67 & 54.00 & 9.50 & 15.50\\
S6 & 69.00 & 74.50 & 78.33 & 81.17 & 50.83 & 59.67 & 9.33 & 15.67\\
S7 & 68.33 & 81.50 & 62.17 & 72.50 & 41.67 & 51.00 & 8.33 & 15.00\\
S8 & 47.17 & 64.17 & 44.83 & 52.00 & 27.83 & 36.83 & 8.83 & 15.83\\
S9 & 70.83 & 79.50 & 70.50 & 78.83 & 48.83 & 58.00 & 9.17 & 15.50\\
S10 & 78.50 & 90.00 & 79.67 & 84.33 & 57.17 & 63.00 & 15.50 & 23.50\\
\bottomrule\end{tabular*}
\end{table*}

With global assignment, the relay improves over direct calibration by 10.92, 6.63, 8.28 and 6.68 points on BHA, SAMGA, Brain-HiVE and ATM-S (\cref{tab:s_backbones}), each decoder predicting its anchors from its own vision layers. Every seed-averaged subject improves on all four decoders (\cref{tab:s_subjects}). Absolute accuracies differ across decoders because their visual features and scores differ.

\subsection{What the gain depends on}
\label{app:components}
\begin{table}[!htb]\centering\footnotesize
\caption{Component deletions: Top-1 (\%), one training seed, ten subjects. Each variant retains image-stage recovery and final CSLS; encoder, anchors, and parameters remain fixed. Bold: best per column.}
\label{tab:s_deletions}
\begin{tabular}{@{}lrr@{}}\toprule
Population-stage intervention & BHA & SAMGA\\\midrule
Without transport initialization & 75.40 & 73.35\\
Without population refinement & 71.30 & 69.70\\
\textbf{Full CHORUS} & \textbf{77.90} & \textbf{73.85}\\
\bottomrule\end{tabular}
\end{table}

\begin{table}[!htb]\centering\footnotesize\setlength{\tabcolsep}{3pt}
\caption{Readout control: Top-1 (\%), one training seed, ten subjects. Cosine and CSLS score identical final embeddings within each route. Bold: higher mean per column.}
\label{tab:s_readout}
\begin{tabular}{@{}lrrrr@{}}\toprule
& \multicolumn{2}{c}{BHA} & \multicolumn{2}{c}{SAMGA}\\
\cmidrule(lr){2-3}\cmidrule(l){4-5}
Calibration & Cosine & CSLS & Cosine & CSLS\\\midrule
Direct & 64.25 & 66.15 & 63.60 & 66.65\\
Population relay & \textbf{75.10} & \textbf{77.90} & \textbf{70.15} & \textbf{73.85}\\
\bottomrule\end{tabular}
\end{table}

\begin{table}[t]
\centering\footnotesize\setlength{\tabcolsep}{2.5pt}
\caption{\textbf{Training schemes under each calibration route} (unlabeled batch, Top-1 \%, three seeds, ten held-out subjects). Each backbone is trained with one method's objective and then calibrated by the same routes: none, direct calibration, and the population relay with anchor predictors fitted and selected on source validation as for SCA.}
\label{tab:s_trainxcal}
\begin{tabularx}{\columnwidth}{@{}l*{6}{Y}@{}}
\toprule
& \multicolumn{3}{c}{BHA} & \multicolumn{3}{c}{SAMGA}\\
\cmidrule(lr){2-4}\cmidrule(l){5-7}
Training & None & Direct & Relay & None & Direct & Relay\\
\midrule
Base & 22.67 & 61.83 & 69.95 & 25.22 & 58.28 & 64.40\\
SATTC~\cite{huang2026sattc} & 23.43 & 61.57 & 71.95 & 25.48 & 55.97 & 65.57\\
SCORE~\cite{cui2026score} & 21.27 & 53.58 & 64.80 & 28.47 & 62.65 & 69.15\\
ProCA~\cite{zhou2026proca} & 24.57 & 62.45 & 74.75 & 25.87 & 58.62 & 65.83\\
\textbf{SCA} & \textbf{28.17} & \textbf{67.45} & \textbf{78.37} & \textbf{36.62} & \textbf{66.77} & \textbf{73.40}\\
\bottomrule
\end{tabularx}
\end{table}

\paragraph{Anchor predictor.}
With anchors predicted from pooled features only, the SCA relay reaches 73.63/71.12\% on BHA/SAMGA, still 6.18/4.35 points above direct calibration for all ten subjects on both backbones (\cref{tab:s_anchorq}).

\paragraph{Calibration components.}
With one training seed, we keep the encoder, anchors and image stage and delete one population-stage component (\cref{tab:s_deletions}). Without transport initialization, refinement starts from the mean-aligned queries; without population refinement, the transport-fitted representation goes directly to the image stage. Removing refinement costs 6.60/4.15 points and lowers accuracy for all ten subjects on both backbones; removing transport costs 2.50/0.50 points, with seven of ten subjects lower on each backbone.

\paragraph{Readout.}
Scoring the same final embeddings with plain cosine instead of CSLS keeps the relay ahead by 10.85/6.55 points, compared with 11.75/7.20 with CSLS (\cref{tab:s_readout}). CSLS improves both routes, but the relay gain comes from the calibration itself.

\paragraph{Training scheme.}
With every training scheme the relay improves over direct calibration, by 8.12 to 12.30 points on BHA and 6.12 to 9.60 on SAMGA, for every held-out subject (\cref{tab:s_trainxcal}); SCA is the best training scheme under each route.

\subsection{Where the prior schemes lose accuracy}
\label{app:rivals}
\begin{table}[!htb]
\centering\footnotesize\setlength{\tabcolsep}{2.5pt}
\caption{\textbf{Prior schemes, training crossed with calibration} (unlabeled batch, Top-1 \%, three seeds, ten held-out subjects). ``Own'': the method's own training objective or calibration.}
\label{tab:s_rivals}
\begin{tabularx}{\columnwidth}{@{}l*{4}{Y}@{}}
\toprule
& \multicolumn{2}{c}{BHA} & \multicolumn{2}{c}{SAMGA}\\
\cmidrule(lr){2-3}\cmidrule(l){4-5}
Training + calibration & SATTC & SCORE & SATTC & SCORE\\
\midrule
Own + own (complete) & 35.23 & 42.05 & 29.85 & 52.47\\
Own + direct & 61.57 & 53.58 & 55.97 & 62.65\\
Own + relay & 71.95 & 64.80 & 65.57 & 69.15\\
SCA + own & 40.15 & 56.48 & 36.87 & 56.52\\
SCA + direct & 67.45 & 67.45 & 66.77 & 66.77\\
\rowcolor{oursbg}
SCA + relay (CHORUS) & \textbf{78.37} & \textbf{78.37} & \textbf{73.40} & \textbf{73.40}\\
\bottomrule
\end{tabularx}
\end{table}

\Cref{tab:s_rivals} crosses the training and the calibration of SATTC and SCORE. Their encoders are sound: with direct calibration they reach 53.58--62.65\%, and with the population relay 64.80--71.95\%. Most of the distance between the complete systems and CHORUS lies in calibration: on each method's own encoder, replacing its calibration by the relay adds 36.72 (SATTC) and 22.75 (SCORE) points on BHA and 35.72 and 16.68 on SAMGA, and replacing its training by SCA then adds 6.42, 13.57, 7.83 and 4.25.

The two calibrations lose accuracy at different steps. SCORE fits its recovery on landmarks found by mutual nearest neighbors on CSLS-corrected similarities; with the same sparse matching, our direct calibration of the SCA encoder reaches 59.78/63.07\% (\cref{app:s_matchers}), so sparse matching accounts for 7.67 of the 10.97 points between SCORE's calibration and direct calibration on BHA and 3.70 of 10.25 on SAMGA. SATTC adapts the query statistics and rescales and re-ranks similarities per query, but infers no correspondences from which to fit a transform; on BHA its calibration of the SCA encoder (40.15\%) stays at the level of CSLS alone (39.17\%; see \cref{app:stages} for the stage readouts), whereas the correspondence-based stages carry the relay from 45.97 to 78.37\%. On the training side, SCORE's objective raises SAMGA's accuracy under every route but lowers BHA's, and SATTC's whitened targets leave both backbones near Base (\cref{tab:s_trainxcal}).

\subsection{Fewer queries, larger galleries and fewer trials}
\label{app:robust}
\begin{table}[!htb]
\centering\footnotesize\setlength{\tabcolsep}{4pt}
\caption{Partial queries and larger galleries: Top-1 (\%), three seeds, ten subjects. Query subsets average three draws; the 1,000-image gallery adds 800 validation-image distractors. D/C: direct/CHORUS; H/M: Hungarian/mutual neighbors (on CSLS-corrected similarities). Bold: higher mean per comparison.}
\label{tab:s_robust}
\begin{tabular}{@{}lrrrrr@{}}\toprule
Queries/gallery & Match & BHA D & BHA C & SAMGA D & SAMGA C\\\midrule
200/200 & H & 67.45 & \textbf{78.37} & 66.77 & \textbf{73.40}\\
200/200 & M & 59.78 & \textbf{69.23} & 63.07 & \textbf{66.92}\\
100/200 & H & 52.41 & \textbf{61.07} & 53.79 & \textbf{59.47}\\
100/200 & M & 51.94 & \textbf{61.22} & 52.60 & \textbf{57.92}\\
50/200 & H & 45.91 & \textbf{53.20} & 45.00 & \textbf{49.64}\\
50/200 & M & 45.93 & \textbf{52.62} & 44.91 & \textbf{49.07}\\
200/1,000 & H & 26.70 & \textbf{33.30} & 29.50 & \textbf{32.25}\\
\bottomrule\end{tabular}
\end{table}

\begin{table}[!htb]
\centering\footnotesize\setlength{\tabcolsep}{4pt}
\caption{Unlabeled batches with fewer target repetitions. Top-1 (\%), averaging three training seeds and ten subjects. Queries average the first $r$ repetitions; anchors and calibration constants remain fixed. Bold: higher mean within each backbone.}
\label{tab:s_repetitions}
\begin{tabular}{@{}lrrrr@{}}\toprule
Trials/image & BHA D & BHA C & SAMGA D & SAMGA C\\\midrule
4 & 18.10 & \textbf{23.52} & 12.20 & \textbf{14.72}\\
10 & 34.83 & \textbf{43.10} & 29.05 & \textbf{32.93}\\
20 & 48.53 & \textbf{58.08} & 44.03 & \textbf{49.02}\\
80 & 67.45 & \textbf{78.37} & 66.77 & \textbf{73.40}\\
\bottomrule\end{tabular}
\end{table}

Both routes see the same query subsets and distractors; each distractor receives a predicted anchor from its image features. With 100 of the 200 queries, the relay leads by 8.66/5.68 points on BHA/SAMGA, with 50 by 7.29/4.64, and in a 1,000-way gallery with 800 validation-image distractors by 6.60/2.75 (\cref{tab:s_robust}). With fewer repetitions per test image, SAMGA relay accuracy is 14.72, 32.93 and 49.02\% at 4, 10 and 20 trials, against 12.20, 29.05 and 44.03\% for direct calibration (\cref{tab:s_repetitions}). Causal streams (63.13/56.36\%, against 53.63/50.93\% for direct calibration) average over all arrivals, including early ones with few repetitions, and therefore stay below the 80-trial batch.

\subsection{Sparse matching}
\label{app:s_matchers}\label{app:s_countcontrol}
\begin{table}[!htb]
\centering\footnotesize\setlength{\tabcolsep}{4pt}
\caption{Stage-specific matcher ablations: Top-1 (\%), three-seed mean $\pm$ SD. H/M: Hungarian/mutual neighbors (on CSLS-corrected similarities); paired letters indicate population/image matchers. D denotes direct calibration. Other settings remain fixed. Bold: best per backbone.}
\label{tab:s_matchers}
\begin{tabular}{@{}lrr@{}}\toprule
Backbone & Stages & Top-1 (\%)\\\midrule
BHA & HH & $\mathbf{78.37}\pm0.90$\\
BHA & MH & $75.90\pm1.26$\\
BHA & HM & $71.68\pm0.29$\\
BHA & MM & $69.23\pm1.78$\\
BHA & DH & $67.45\pm1.19$\\
BHA & DM & $59.78\pm1.08$\\
SAMGA & HH & $\mathbf{73.40}\pm0.61$\\
SAMGA & MH & $71.05\pm0.95$\\
SAMGA & HM & $69.58\pm0.55$\\
SAMGA & MM & $66.92\pm1.11$\\
SAMGA & DH & $66.77\pm0.25$\\
SAMGA & DM & $63.07\pm0.51$\\
\bottomrule\end{tabular}
\end{table}

\begin{table}[!htb]
\centering\small
\setlength{\tabcolsep}{3.8pt}
\caption{Matched-count control (SAMGA, one training seed, ten subjects). H/M: Hungarian/mutual neighbors (on CSLS-corrected similarities). Random subsets average 20 draws. Count/precision are subject means; Image H/M gives final Top-1 (\%). Bold: best retrieval per column.}
\label{tab:s_countcontrol}
\begin{tabular}{lrrrr}\toprule
Population pairs & Count & Prec. (\%) & Image H & Image M\\\midrule
All H pairs & 200.0 & 74.05 & \textbf{73.85} & \textbf{70.10}\\
Random H subset & 133.7 & 74.21 & 73.36 & 69.42\\
H on M query rows & 133.7 & 82.65 & 72.55 & 68.65\\
Original M pairs & 133.7 & 79.17 & 72.10 & 68.15\\
\bottomrule\end{tabular}
\end{table}

Hungarian assignment (H) keeps a full one-to-one assignment; mutual nearest neighbors (M) keep only pairs that are nearest in both directions. As in SCORE, M is selected on CSLS-corrected similarities; on raw cosine a few candidates are the nearest neighbor of many queries and each can keep at most one pair. Hungarian assignment is unchanged by CSLS, whose row and column terms add the same constant to every one-to-one assignment. Two letters give the population- and image-stage matcher, and D denotes direct calibration. With M in both stages, the relay gains 9.45 points on BHA, 3.85 on SAMGA, 7.02 on Brain-HiVE and 6.05 on ATM-S over direct calibration with the same matcher, for all ten subjects on every decoder (\cref{tab:s_backbones}). With M, direct calibration itself (59.78/63.07\%) stays above SCORE's calibration of the same SCA encoder (56.48/56.52\%; main Tab.~2), for all ten subjects and in every training seed.

On SAMGA, replacing only the population matcher costs 2.35 points (73.40 to 71.05\%) and only the image matcher 3.82 (to 69.58\%); replacing both costs 6.48 (to 66.92\%), whereas direct calibration loses 3.70 (66.77 to 63.07\%; \cref{tab:s_matchers}). In the population stage, H keeps 200 pairs at 73.45\% precision (146.90 correct) and M keeps 133.80 at 78.07\% (105.23 correct): sparse matching drops more than a quarter of the correct constraints for a modest gain in precision.

To separate how many pairs from which pairs, we refit the population transform on SAMGA (one training seed) from frozen pair sets (\cref{tab:s_countcontrol}): all 200 H pairs, random H subsets of the M count (20 draws), H pairs restricted to the query rows that M kept, and the original M pairs. Subsampling H to 133.7 pairs costs only 0.5 points (73.85 to 73.36\%), whereas the M pairs reach 72.10\%. H restricted to M's rows reaches 72.55\% despite higher precision (82.65 versus 74.05\%). The loss under sparse matching thus comes from both fewer pairs and which pairs remain, and precision alone does not predict calibration quality.

\subsection{Alljoined-1.6M}
\label{app:xdataset}\label{app:alljoined_stream}
\begin{table}[!htb]
\centering\footnotesize\setlength{\tabcolsep}{4pt}
\caption{Alljoined-1.6M: 200-way Top-1 (\%), one training seed and 20 held-out participants. Both settings use the predicted anchors; streams average three arrival orders. Bold: higher mean within each setting.}
\label{tab:s_alljoined}
\begin{tabular}{@{}lrrrr@{}}\toprule
Encoder & Setting & Direct & CHORUS & $\Delta$\\\midrule
Base & Unlabeled batch & 15.48 & \textbf{16.88} & +1.40\\
Base & Simulated stream & 11.20 & \textbf{12.40} & +1.20\\
Base + SCA & Unlabeled batch & 18.20 & \textbf{20.23} & +2.03\\
Base + SCA & Simulated stream & 12.89 & \textbf{14.36} & +1.47\\
\bottomrule\end{tabular}
\end{table}

\begin{figure*}[t]
\centering
\includegraphics[width=.94\textwidth]{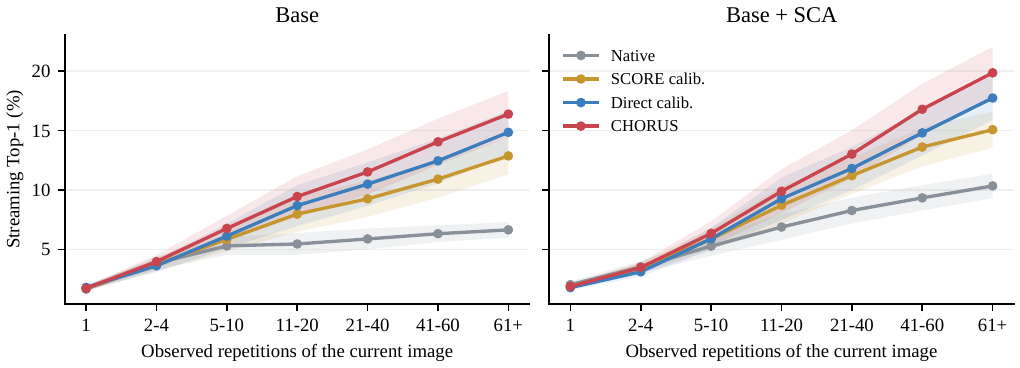}
\caption{\textbf{Alljoined simulated stream} by the number of repetitions observed for the current image (one training seed, 20 subjects). Each point averages three orders, then subjects; shading: subject SEM. All methods share inputs and arrival schedule.}
\label{fig:s_alljoined_stream}
\end{figure*}
Alljoined-1.6M~\cite{alljoined2025} records 20 participants with a consumer-grade 32-channel wet-electrode system at 250\,Hz. We use the released preprocessing, the same concept split and 200 test candidates, and train BHA on the other 19 participants in each fold (one training seed); without calibration, the Base and SCA encoders reach 6.53 and 10.50\%. Only unlabeled calibration is evaluated. Patch-token residuals use BHA's DINOv2-L layer 12, MAE layer 18 and DFN layer 6; the OpenCLIP features used in training were extracted with random crops and enter only the pooled ridge. Every fold selects a residual configuration on source validation.

In the batch, the relay reaches 16.88 versus 15.48\% on Base and 20.23 versus 18.20\% on SCA, where 13 of 20 participants favor the relay and four direct calibration; SCORE calibration of the same SCA encoder reaches 15.20\% (\cref{tab:s_alljoined}). The simulated stream uses the retained single trials in three orders, randomized within blocks of repetitions 1--20, 21--40, 41--60 and the rest, and the procedure of \cref{app:s_stream_steps}; SCORE refits its recovery on the same history. The relay reaches 12.40/14.36\% on Base/SCA, 1.20/1.47 points above direct calibration, with 18 and 19 of 20 participants improving. \Cref{fig:s_alljoined_stream} breaks the stream down by the number of repetitions seen so far.

\section{Additional analyses}
\label{app:analysis}
All analyses in this section use BHA with three training seeds and ten held-out subjects unless stated otherwise.

\subsection{Geometry and shared residuals}
\label{app:geometry}
\begin{table}[!htb]\centering\footnotesize
\caption{Agreement of each reference with the held-out subject (BHA, 30 folds). The measured population uses source test EEG (offline). Cosine: per-image agreement after centering each set.}
\label{tab:s_geometry}
\begin{tabular}{@{}lrrr@{}}\toprule
Reference & CKA & 10-NN & Centered cosine\\\midrule
Images & .590 & .311 & .361\\
Predicted population & .673 & .340 & .495\\
Measured population & .716 & .395 & .553\\\bottomrule
\end{tabular}
\end{table}
Each embedding set is centered and row-normalized; we report unbiased linear CKA~\cite{kornblith2019similarity}, mutual 10-nearest-neighbor overlap and the per-image centered cosine (\cref{tab:s_geometry}). The measured population agrees with the held-out subject more than the best single source (CKA 0.72 versus 0.62) and an average single source (0.58), and predicted anchors keep part of this advantage globally and locally. The held-out subject's EEG-minus-image residual has a mean cosine of 0.58 with the population's residual and 0.56 between two sources; predicted and measured population residuals reach 0.80, and shuffling the stimulus pairing gives 0.00. The shared encoder, common visual supervision, and subtraction of the same image embedding can all contribute to this shared deviation; it is not by itself evidence of a uniquely neural signal.

After centering, row normalization and optimal orthogonal alignment, the held-out subject is still closer to the population than to the images for 85\% of concepts over all subjects (main Fig.~6c; the figure uses the 30 subject--seed folds and paired subject tests).

Across the ten seed-averaged subjects, the local-agreement advantage of the predicted anchors over the images correlates with final CHORUS accuracy (Spearman $\rho=.66$, $p=.039$), whereas agreement with the images alone does not ($\rho=.21$). The advantage does not correlate with the gain over direct calibration ($\rho=-.23$), so it describes which subjects are easy rather than how much the relay adds. Main Fig.~5d instead relates agreement to the correspondences the calibrator finds; \cref{app:anat_pairs} repeats this on four decoders.

\subsection{Calibration stages and prediction coverage}
\label{app:stages}
\begin{figure*}[t]\centering
\includegraphics[width=\textwidth]{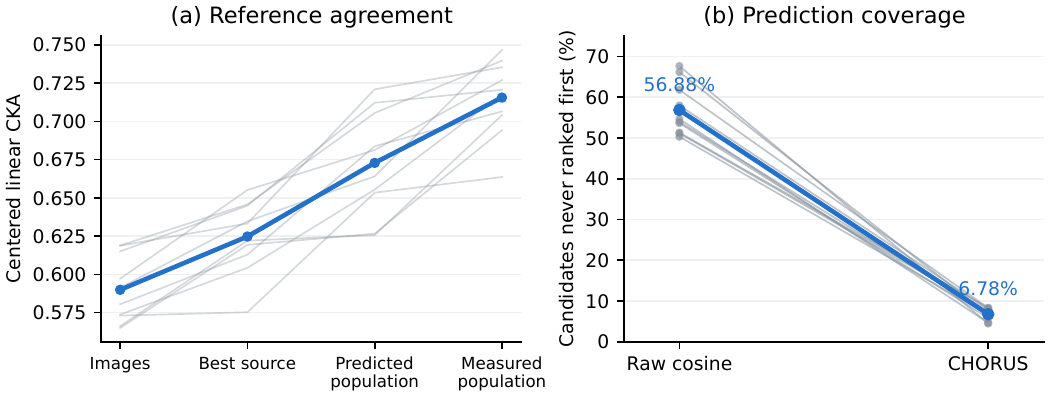}
\caption{\textbf{Reference agreement and prediction coverage} (BHA). Thin/thick lines: seed-averaged subject/overall means. (a) Agreement of each reference with the held-out subject; best source and measured population are offline references. (b) Fraction of gallery images never retrieved first, from uncalibrated cosine to CHORUS with CSLS.}
\label{fig:s_reference_coverage}\end{figure*}
\begin{table}[!htb]\centering\small
\caption{BHA calibration stages: three seeds, ten subjects. Every representation is evaluated against the same 200-image gallery with CSLS; deployment uses only the final representation.}
\label{tab:s_stage_readout}
\begin{tabular}{lr}\toprule
Intermediate representation & Image Top-1 (\%)\\\midrule
Uncalibrated & 39.17\\
Source-mean alignment & 45.97\\
Transport initialization and rotation & 62.48\\
Population assignment refinement & 69.48\\
Image-stage recovery & 78.37\\
\midrule
Direct image calibration & 67.45\\
\bottomrule\end{tabular}
\end{table}

Scoring every intermediate representation against the same gallery with CSLS (\cref{tab:s_stage_readout}) shows how the two stages divide the work. Accuracy rises from 39.17\% (28.17\% with native cosine) to 69.48\% after population refinement and to 78.37\% after the image stage, against 67.45\% for direct calibration. The population-aligned representation is not yet the best image readout, but it is a better starting point for the image stage. Calibration also spreads the predictions: the fraction of gallery images never ranked first falls from 56.88 to 6.78\%, and the largest number of queries choosing one image from 14.4 to 2.4 (\cref{fig:s_reference_coverage}b). Hungarian assignment is used only to fit the transforms; final predictions are independent row maxima.

\subsection{Anchor quality}
\label{app:interp}
\begin{table}[!htb]
\centering\footnotesize\setlength{\tabcolsep}{4pt}
\caption{Offline interpolation from predicted ($\alpha=0$) to measured ($\alpha=1$) population anchors, BHA. Measured anchors use source test EEG. Cosine is centered agreement with the measured reference. All target embeddings and calibration settings are fixed.}
\label{tab:s_interpolate}
\begin{tabular}{@{}lrr@{}}\toprule
$\alpha$ & Anchor cosine & Top-1 (\%)\\\midrule
0 & 0.871 & 78.37\\
0.25 & 0.938 & 83.53\\
0.5 & 0.977 & 85.33\\
0.75 & 0.995 & 85.33\\
1 & 1.000 & 85.00\\
\bottomrule\end{tabular}
\end{table}

We interpolate $A_\alpha=(1-\alpha)\widehat A+\alpha A_{\rm measured}$ while keeping target embeddings, gallery and calibration fixed (\cref{tab:s_interpolate}, main Fig.~7b). Accuracy rises from 78.37\% with predicted anchors to 85.33\% at $\alpha=.5$ and $.75$ and is 85.00\% at the measured endpoint. Measured anchors require source test EEG, so this is an offline estimate of how much a better predictor could add.

\subsection{Source count and equal trial budgets}
\label{app:s_diversity}\label{app:s_equaltrials}
\begin{table}[!htb]
\centering\footnotesize\setlength{\tabcolsep}{4pt}
\caption{BHA source-count analysis: Top-1 (\%), three-seed mean $\pm$ SD. Three subsets per size except nine sources. At $K=200$, the subset supplies both the measured support anchors and the predicted gallery anchors; five support draws are averaged. Direct: 67.45\% (batch), 54.08\% ($K=200$).}
\label{tab:s_count}
\begin{tabular}{@{}lrr@{}}\toprule
Sources & Unlabeled batch & $K=200$\\\midrule
1 & $71.69\pm1.34$ & $45.13\pm0.48$\\
2 & $73.40\pm1.43$ & $52.78\pm0.59$\\
3 & $75.27\pm0.97$ & $55.60\pm0.67$\\
5 & $77.02\pm0.80$ & $59.05\pm0.85$\\
7 & $77.74\pm0.99$ & $60.23\pm0.67$\\
9 & $78.37\pm0.90$ & $60.41\pm0.73$\\
\bottomrule\end{tabular}
\end{table}

To vary the population, we rebuild everything that depends on it from each source subset: regression targets, source means and predicted anchors for the batch, and additionally the measured support anchors for $K=200$, where only the ridge penalty is reselected on the subset's validation images. The encoder remains the one trained on all nine sources (\cref{tab:s_count}). With one source, predicted anchors already give 71.69\%, above direct calibration (67.45\%); nine sources reach 78.37\% (all ten subjects improve). With $K=200$ labels the dependence is stronger, because both the support anchors and the gallery anchors come from the subset: one source gives 45.13\%, below direct calibration (54.08\%) for every subject; two are on par ($-1.30$ points), three exceed it ($+1.52$), from five on every subject favors the relay, and nine reach 60.41\%.

At a fixed total of 80 source trials per image (main Fig.~7d), measured anchors from one source with 80 repetitions give 76.51\%, and from eight sources with ten repetitions each 83.19\%, a gain of 6.68 points with every subject improving. Each source averages its selected repetitions before encoding, and each condition averages three source draws per target. Averaging across people therefore helps beyond averaging more recordings of one person.

\subsection{Transform family, rotation rank and held-out concepts}
\label{app:half}
\begin{table}[!htb]
\centering\footnotesize\setlength{\tabcolsep}{4pt}
\caption{Half-session population identification (100-way): mean $\pm$ seed SD over three seeds, ten subjects, and both split directions. Residual regularization uses nested five-fold validation within the fitting concepts.}
\label{tab:s_half}
\begin{tabular}{@{}lrr@{}}\toprule
Labels $K$ & Transform & Identification (\%)\\\midrule
20 & Rotation & $76.67\pm0.63$\\
50 & Rotation & $84.70\pm0.28$\\
100 & Rotation & $87.42\pm0.34$\\
20 & + residual ridge & $76.33\pm0.49$\\
50 & + residual ridge & $83.98\pm0.19$\\
100 & + residual ridge & $88.18\pm0.20$\\
\bottomrule\end{tabular}
\end{table}

These diagnostics fit a map from the held-out subject to the measured population on some test concepts and identify the others.

\begin{figure}[!htb]
\centering
\includegraphics[width=\columnwidth]{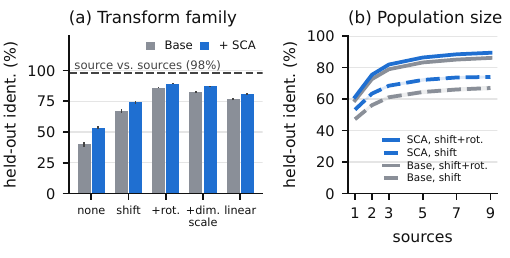}
\caption{\textbf{What calibration corrects} (BHA; offline diagnostic). (a) Held-out 100-way population identification in a 60-dimensional PCA space; dim.\ scale: per-dimension scaling. Dashed: source-subject reference. (b) Varying source count. Bars/bands: subject SEM after averaging three seeds.}
\label{fig:mech}
\end{figure}
\paragraph{Transform family (\cref{fig:mech}).}
Using all 80 repetitions, we fit on one half of the 200 test concepts and identify the other half (100-way), in both directions, in a 60-dimensional PCA space fitted on the fitting concepts. Evaluation queries are centered with their own batch mean and the population with the fitting-set mean. Without a transform, identification is 53.7\%; a mean shift gives 74.1\%, shift plus rotation 89.4\%, adding per-coordinate scaling 87.2\% and a general linear map 80.9\%, against 97.7\% for one source subject identified against the others. With one source as the reference, shift plus rotation reaches only 61.3\%, rising steadily to 89.4\% with nine (\cref{fig:mech}b).

\paragraph{Rotation rank.}
Keeping the full 1,024 dimensions and rotating only within a fitted principal subspace (identity elsewhere; five random half splits), identification reaches 90.32 and 91.56\% at ranks 32 and 64. At rank 64, held-out image retrieval after aligning to the population is 66.45\%, against 59.97\% after aligning to the images.

\paragraph{Half-session transfer.}
To test transfer with independent recordings, we split the 80 repetitions into even and odd halves of 40, use the even half as queries, fit on $K=20$, 50 or 100 concepts and identify the remaining 100 (\cref{tab:s_half}). A shift and a regularized rotation ($\rho=.1$) reach 76.67, 84.70 and 87.42\%. Adding a ridge correction on the rotated residuals, with its strength chosen by five-fold validation inside the fitting concepts, helps only at $K=100$ (88.18\%). Without labels, a relay fitted on all 100 fitting responses reaches 84.40\% with predicted and 87.05\% with measured anchors; the subject's own odd half identifies its even half at 99.90\%.

\begin{table}[!htb]
\centering\footnotesize\setlength{\tabcolsep}{4pt}
\caption{Frozen calibration on new images (100-way Top-1, \%). Each route is fitted without labels on one half of the test concepts, frozen and applied to the other half; mean over ten subjects, five splits and both directions. Last two rows: fitted on the evaluated half itself.}
\label{tab:s_frozen}
\begin{tabular}{@{}lrr@{}}\toprule
Calibration of the new half & BHA & SAMGA\\\midrule
None & 50.06 & 55.36\\
Shift only & 57.13 & 56.75\\
Direct calibration, frozen & 71.98 & 72.35\\
\rowcolor{oursbg}CHORUS, frozen & \textbf{75.13} & \textbf{74.72}\\\midrule
Direct calibration, fitted on this half & 74.53 & 76.87\\
\rowcolor{oursbg}CHORUS, fitted on this half & \textbf{85.98} & \textbf{85.16}\\
\bottomrule\end{tabular}
\end{table}

\paragraph{Frozen calibration on new images.}
To test whether a calibration fitted without labels carries over to images it has not seen, we split the 200 test concepts into two halves of 100 (five random splits, both directions), fit CHORUS and direct calibration on the queries of one half, freeze every fitted quantity (shifts, rotations, moment matching and recovery), apply them to the queries of the other half, and retrieve among its 100 images with the main readout (\cref{tab:s_frozen}; one training seed). Main Table~5 also varies which stage transfers. With the population stage frozen, we pass the evaluation queries through its original shift and rotation and refit the image stage on those outputs and the new gallery. With the image stage frozen, we refit the population stage on the evaluation queries and their candidate anchors, then apply the original image transform. Both conditions use the same batch CSLS readout as the fully frozen maps. The frozen relay leads frozen direct calibration by 3.15 points on BHA and 2.37 on SAMGA, for every held-out subject on both backbones ($p=.002$). The lead holds when the new queries are re-centered with their own mean ($+3.23$ and $+1.96$) and with a plain cosine readout ($+3.73$ and $+2.20$). On the fitting half, 85.9 and 85.3\% of the relay's image-stage pairs are correct, against 74.0 and 76.9\% for direct calibration: the population stage leaves the new subject with a better map, not only with better pairs for the batch at hand.

\paragraph{How many unlabeled images are needed for transfer?}
We vary the number of distinct calibration images, $K$, while keeping the 100 evaluation images fixed within each split. From the other half of the concepts, we take nested subsets of $K$ images and their EEG responses. The EEG, image and predicted-anchor banks contain the same $K$ stimuli in independently shuffled orders. Both routes infer correspondences on this support alone. We then freeze the maps and rank the 100 held-out images by cosine for each query independently. At $K=0$ we score the original embedding. Each response averages 80 repetitions of one stimulus.

\begin{figure}[!htb]
\centering
\includegraphics[width=\columnwidth]{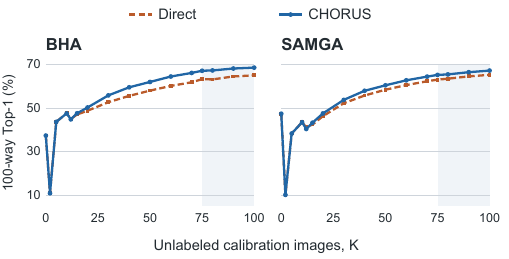}
\caption{\textbf{Transfer from $K$ unlabeled calibration images.} Held-out Top-1 averaged over ten subjects, three encoder seeds and five random half splits in both directions. Shading marks $K\geq75$, where CHORUS stays within two percentage points of its $K=100$ accuracy on both encoders.}
\label{fig:s_holdout_k}
\end{figure}
Two calibration images give worse held-out retrieval than leaving the encoder uncalibrated on both backbones (\cref{fig:s_holdout_k}). With 50 images, the population relay leads direct calibration by 3.97 points on BHA and 2.01 on SAMGA. At $K=75$, the relay reaches 67.16\% and 65.26\%, respectively, against 68.60\% and 67.24\% at $K=100$. Thus an unlabeled batch can support calibration for subsequent images, but a handful of stimuli gives an unreliable estimate of the map.

\subsection{What SCA changes}
\label{app:s_layerwise}
\begin{figure}[!htb]\centering
\includegraphics[width=\columnwidth]{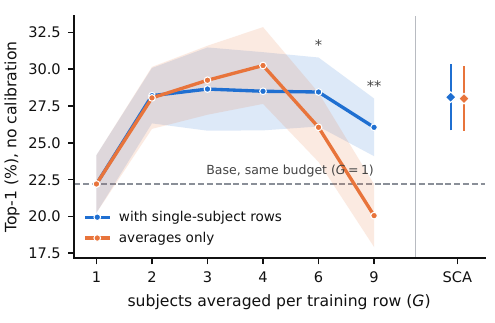}
\caption{\textbf{Fixed-update SCA control} (BHA, one training seed, ten subjects). Blue: individual and averaged rows; orange: averages only. Bands: subject SEM; stars: paired Wilcoxon tests ($^{*}p<.05$, $^{**}p<.01$).}
\label{fig:s_sca}\end{figure}
\begin{figure*}[t]\centering
\includegraphics[width=\textwidth]{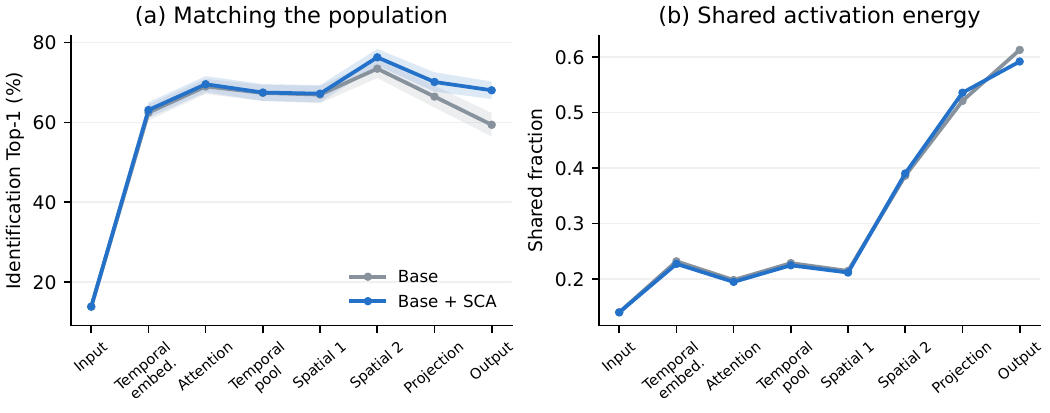}
\caption{\textbf{Layer-wise BHA representations.} (a) Centered held-out-to-population identification. (b) Fraction of activation energy in the across-subject mean. Bands: SEM across ten subjects after averaging three seeds.}
\label{fig:s_layerwise}\end{figure*}

\paragraph{Correspondence, not global similarity (main Fig.~8).}
SCA raises 200-way identification of the held-out subject against the population from 33.82 to 47.22\% and against one source from 22.54 to 29.74\%, and widens the margin between the correct and the best wrong population image from .016 to .043. Global similarity does not increase: the share of source embedding energy in their mean falls from .604 to .587, and the per-image cosine between sources from .591 to .568. With predicted anchors, SCA raises population-stage pair accuracy from 66.32 to 77.25\% and image-stage pair accuracy from 68.78 to 78.23\%.

\paragraph{Fixed number of updates.}
To rule out extra exposure, a control with one training seed trains every arm on the same $2\times2.75n$ rows ($n$: individual rows) and evaluates the final checkpoint (\cref{fig:s_sca}). Keeping individual rows and averaging two to six sources in the rest raises native accuracy from 22.20 to 28.20--28.65\%; nine-person averages give 26.05\% with individual rows and 20.05\% without them.

\paragraph{Where in the encoder.}
With the Base and SCA encoders frozen, we center each subject's activations for the 200 test images at every stage and identify the held-out subject against the nine-source average (\cref{fig:s_layerwise}). Early stages change by less than 0.7 points; identification rises from 73.45 to 76.30\% at the second spatial layer, from 66.45 to 70.12\% at the projection, and from 59.38 to 68.03\% at the output. The share of energy in the across-subject mean at the output falls from .613 to .592, so SCA improves image-specific matching across people without making activations more alike overall.

\section{The calibrator across four decoders}
\label{app:anatomy}
This section takes the unlabeled two-stage calibrator apart on all four decoders: BHA, SAMGA, Brain-HiVE and ATM-S, with the encoders and patch-token anchors of \cref{tab:s_backbones}. Each analysis replays the deployed calibrator on one training seed and ten held-out subjects, so values differ slightly from the three-seed averages of \cref{app:decoders}; nothing is refitted or tuned. Measured anchors (the source-population average of the test images) need source test EEG and appear only as an offline reference.

\subsection{Stage by stage}
\label{app:anat_stages}
\begin{figure*}[t]\centering
\includegraphics[width=\textwidth]{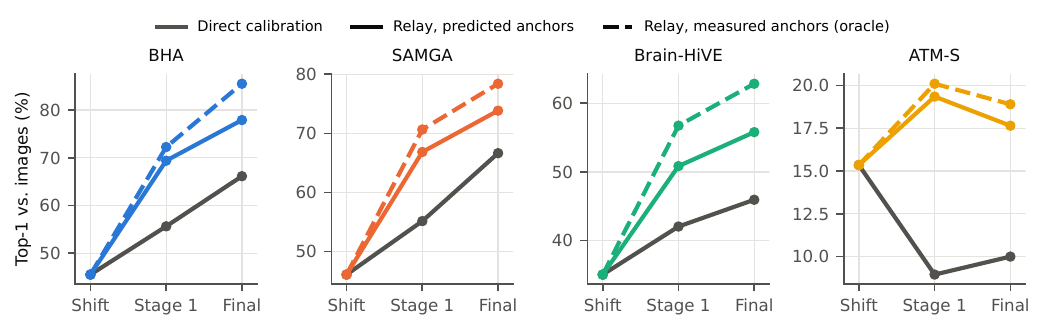}
\caption{\textbf{Image Top-1 after each stage.} Shift: mean alignment only; Stage 1: after the first step (toward the images for direct calibration, toward the anchors for the relay); Final: complete calibration. One training seed, ten held-out subjects.}
\label{fig:s_anat_stages}\end{figure*}

\begin{table*}[t]\centering\footnotesize\setlength{\tabcolsep}{5pt}
\caption{\textbf{Stage readouts and pair precision} (\%). Every representation is scored against the same 200-image gallery with CSLS unless stated; pair precision is the fraction of correct correspondences used to fit each stage. One training seed, ten held-out subjects.}
\label{tab:s_anat}
\begin{tabular}{@{}lrrrr@{}}\toprule
 & BHA & SAMGA & Brain-HiVE & ATM-S\\\midrule
Mean shift only & 45.55 & 46.10 & 35.05 & 15.35\\
Direct calibration, after the first step & 55.65 & 55.15 & 42.05 & 8.95\\
Direct calibration & 66.15 & 66.65 & 45.95 & 10.00\\
Relay, after the population stage & 69.40 & 66.85 & 50.85 & 19.35\\
Relay, after the population stage, read against predicted anchors & 72.60 & 67.30 & 54.45 & 14.75\\
\textbf{Relay (CHORUS)} & \textbf{77.90} & \textbf{73.85} & \textbf{55.80} & \textbf{17.65}\\
Relay with measured anchors (offline) & 85.50 & 78.40 & 62.85 & 18.90\\\midrule
Correct initial pairs: images / predicted / measured anchors & 67.5/73.2/90.6 & 68.3/73.6/90.9 & 51.5/58.0/83.8 & 20.2/30.3/78.1\\
Correct image-stage pairs: direct / relay & 65.3/77.8 & 65.2/74.5 & 43.9/54.9 & 10.1/21.7\\\midrule
Mutual NN in both stages: direct / relay & 58.60/67.40 & 63.65/68.15 & 43.80/50.35 & 8.85/15.55\\\bottomrule
\end{tabular}
\end{table*}

The population stage gives the image stage a better start on every decoder (\cref{fig:s_anat_stages,tab:s_anat}). After it, image Top-1 is 13.75, 11.70, 8.80 and 10.40 points above the first step of direct calibration, and the image stage then fits on 12.45, 9.30, 11.05 and 11.65 points more correct pairs (all ten subjects on every decoder). The relay ends 11.75, 7.20, 9.85 and 7.65 points above direct calibration, again for all ten subjects. Read out against the predicted anchors right after the population stage, the relay is already above complete direct calibration on BHA, Brain-HiVE and ATM-S ($+6.45$, $+8.50$, $+4.75$; ten of ten subjects) and on par on SAMGA; the image stage then adds the rest, so the anchors serve as a relay toward the images rather than as the final gallery. With mutual nearest neighbors in both stages (selected on CSLS-corrected similarities, as in SCORE), the relay again improves over direct calibration with the same matcher (BHA $+8.80$, SAMGA $+4.50$, Brain-HiVE $+6.55$, ATM-S $+6.70$), and measured anchors raise it further ($+7.15$ to $+14.85$ over direct calibration, ten of ten subjects).

\subsection{The two gaps in each decoder's space}
\label{app:anat_gaps}
\begin{figure*}[t]\centering
\includegraphics[width=\textwidth]{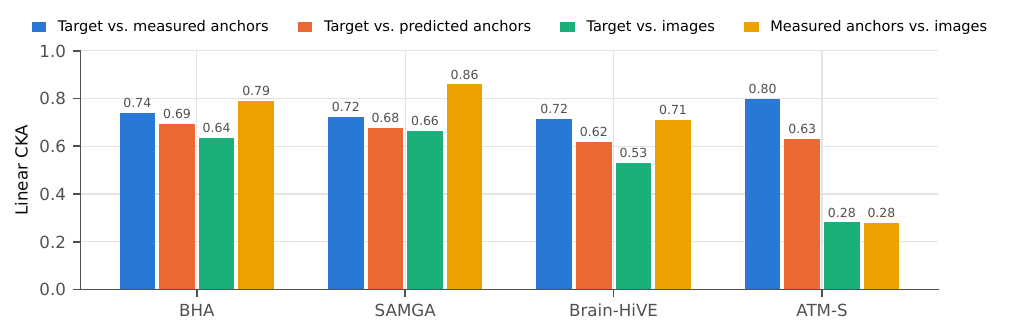}
\caption{\textbf{The two gaps in each decoder's own embedding space} (linear CKA on the 200 test images; higher is closer). Blue: new subject vs.\ population (measured anchors); green: new subject vs.\ images; yellow: population vs.\ images; orange: new subject vs.\ predicted anchors.}
\label{fig:s_anat_gaps}\end{figure*}

\begin{table}[!htb]\centering\footnotesize\setlength{\tabcolsep}{3pt}
\caption{\textbf{Encoder geometry.} RSA: Spearman correlation of centered cosine-similarity matrices. Identification: 200-way Top-1 (\%) of the mean-shifted target among measured anchors or images. Effective dimension: participation ratio.}
\label{tab:s_anat_geom}
\begin{tabular}{@{}lrrrr@{}}\toprule
 & BHA & SAMGA & B-HiVE & ATM-S\\\midrule
RSA target--measured & .70 & .68 & .68 & .76\\
RSA target--predicted & .66 & .64 & .57 & .57\\
RSA target--images & .56 & .62 & .46 & .18\\
CKA measured--images & .79 & .86 & .71 & .28\\
CKA predicted--images & .84 & .96 & .88 & .27\\\midrule
Identification among measured & 72.9 & 69.5 & 61.1 & 55.1\\
Identification among images & 45.6 & 46.1 & 35.1 & 15.4\\\midrule
Effective dim.: target / images & 17/36 & 16/11 & 13/37 & 9/68\\\bottomrule
\end{tabular}
\end{table}

In every decoder, a new subject is closer to the population than to the images (\cref{fig:s_anat_gaps,tab:s_anat_geom}): before any calibration, its mean-shifted responses identify 55--73\% of the test images among the measured anchors but 15--46\% among the images, and its RSA with the population exceeds its RSA with the images by .13, .06, .22 and .58 (ten of ten subjects on every decoder). The new subject--population gap is similar across decoders (CKA .72--.80), whereas the population--image gap differs widely: SAMGA's EEG space is closest to its image space (.86), ATM-S's farthest (.28). The predicted anchors lie between population and images and lean toward the images (CKA .84--.96 with the images on BHA, SAMGA and Brain-HiVE), because they are regressed from visual features.

\subsection{Agreement decides the correspondences}
\label{app:anat_pairs}
\begin{figure}[!htb]\centering
\includegraphics[width=\columnwidth]{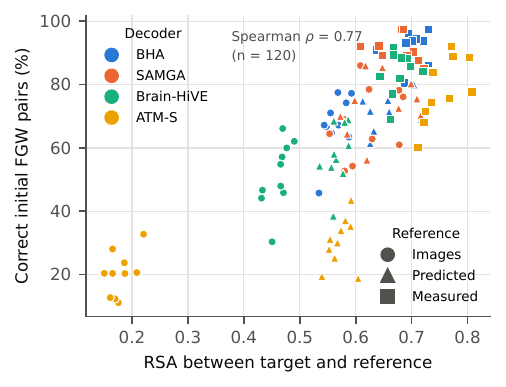}
\caption{\textbf{Structural agreement predicts correct correspondences.} One point per decoder, reference and held-out subject; circles: images (direct calibration), triangles: predicted anchors, squares: measured anchors.}
\label{fig:s_anat_rsa}\end{figure}

Transport initializes only the first stage, combining cross-set cosine distances with within-set geometric consistency. The image stage starts from population-aligned embeddings and infers fresh Hungarian assignments. The more a reference shares the target's structure, the more of the transport proposals are correct (\cref{fig:s_anat_rsa}): across decoders, references and subjects, RSA with the target predicts the fraction of correct initial pairs (Spearman $\rho=.77$, $n=120$; .44--.87 within each decoder). On every decoder the order is the same, measured anchors above predicted anchors above images, and the predicted anchors give 5.7, 5.3, 6.4 and 10.0 points more correct initial pairs than the images. The population is thus the easier reference to align to without labels.

\subsection{Labeled support, query by query}
\label{app:anat_labeled}
\begin{figure*}[t]\centering
\includegraphics[width=\textwidth]{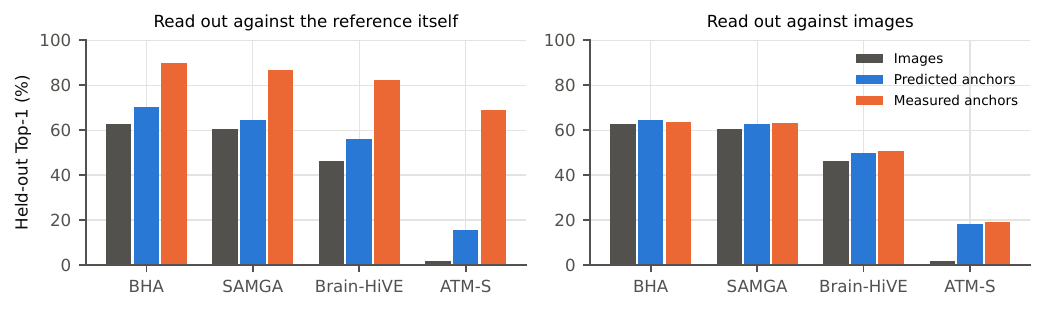}
\caption{\textbf{Labeled rotation, each query scored on its own.} The calibrator's regularized rotation is fitted on 100 true pairs toward each reference; the other 100 queries are scored individually among 200 candidates (two folds). Left: scored against the reference itself, as with labeled support in the main paper for the predicted anchors; right: scored against the images.}
\label{fig:s_anat_labeled}\end{figure*}

The same advantage holds when correspondences are known. Fitting the rotation on the same true pairs and scoring each held-out query on its own, the relay (fit to and scored against the predicted anchors) exceeds direct calibration (fit to and scored against the images) by 7.50, 4.25, 9.85 and 13.95 points (ten, nine, ten and ten of ten subjects; \cref{fig:s_anat_labeled}), and still by 1.65, 2.55, 3.45 and 16.65 points when the relay is also scored against the images. Measured anchors add nothing significant over predicted ones here. Because the target is closer to the population, less of its residual must be explained by the fitted pairs, and the rotation generalizes better to held-out queries; this is the setting of the labeled-support results in the main paper, where the relay leads at every budget.

\subsection{Where calibrated queries land}
\label{app:anat_where}
Both first steps are a shift and a rotation, which leave the target's internal distances unchanged. Plots of the target embeddings alone (PCA, t-SNE or CKA) therefore look the same after either calibration; what changes is where each query lands relative to the gallery.

\begin{figure*}[t]\centering
\includegraphics[width=\textwidth]{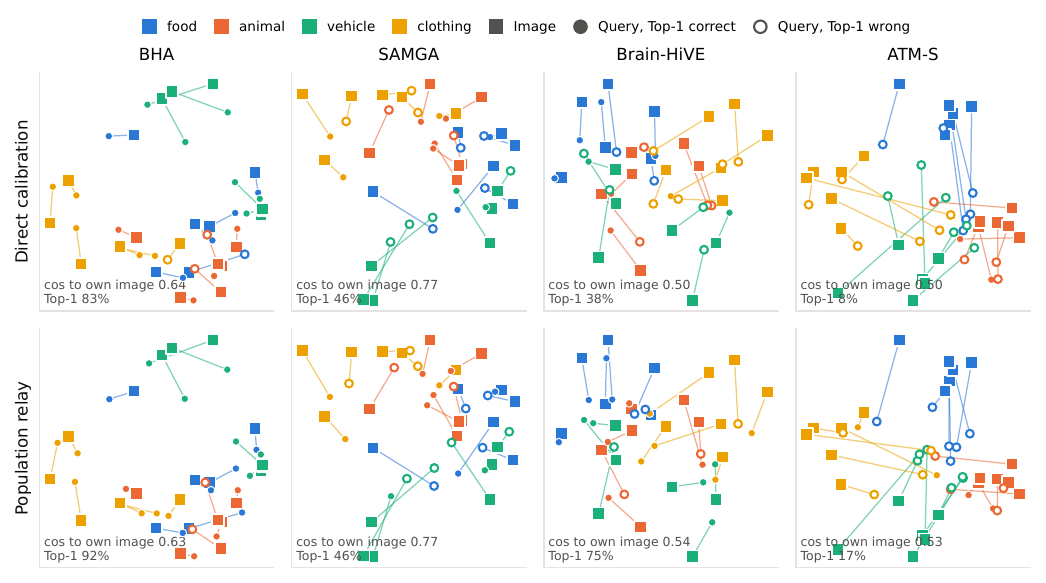}
\caption{\textbf{Joint projection} (illustration). For a median-gain subject per decoder, six single-category images from each of four THINGS categories (lowest image indices) and their calibrated queries, on the first two principal components of these image embeddings; lines join each query to its own image; filled: 200-way Top-1 correct. Top: direct calibration; bottom: relay.}
\label{fig:s_anat_proj}\end{figure*}

\begin{figure*}[t]\centering
\includegraphics[width=\textwidth]{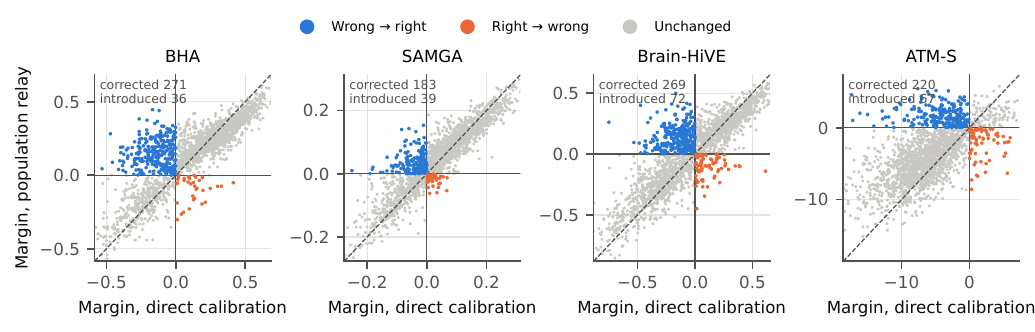}
\caption{\textbf{Per-query margin}: CSLS score of the true image minus the best competitor ($>0$: correct), direct calibration (x) vs.\ relay (y), 2{,}000 queries per decoder. Blue: wrong to right; orange: right to wrong. ATM-S uses its native dot product.}
\label{fig:s_anat_margin}\end{figure*}

\begin{figure*}[t]\centering
\includegraphics[width=\textwidth]{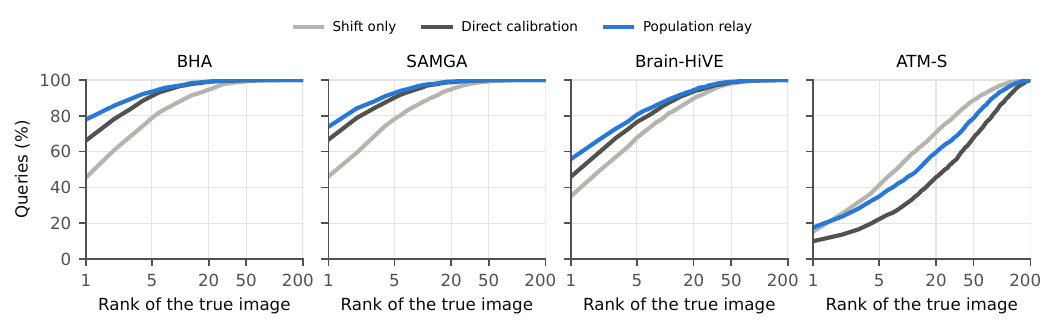}
\caption{\textbf{True-image rank.} Each curve shows the percentage of 2{,}000 queries per decoder retrieved in the top $k$ ($k=1$--$200$, log scale).}
\label{fig:s_anat_rank}\end{figure*}

The relay does not move queries closer to their own images on average (mean cosine 0.606 vs.\ 0.604 on BHA, 0.767 vs.\ 0.770 on SAMGA, 0.500 vs.\ 0.516 on Brain-HiVE, 0.493 vs.\ 0.517 on ATM-S for direct calibration vs.\ relay); it changes which image comes first (\cref{fig:s_anat_proj}). The per-query margins show where (\cref{fig:s_anat_margin}): many more queries go from wrong to right than from right to wrong (271 vs.\ 36 on BHA, 183 vs.\ 39 on SAMGA, 269 vs.\ 72 on Brain-HiVE, 220 vs.\ 67 on ATM-S), almost all of them narrowly wrong under direct calibration, and the mean margin rises for ten, nine, ten and ten of ten subjects. The whole rank distribution shifts forward (\cref{fig:s_anat_rank}): the true image is in the top five for 93.5, 92.7, 80.6 and 35.0\% of queries with the relay, against 91.1, 90.1, 76.5 and 22.3\% with direct calibration. The remaining errors are as semantic under both routes: 29--40\% of wrong first choices share a THINGS category with the true image, against 14.4\% by chance.

\subsection{How the gain divides between the stages}
\label{app:anat_decoders}
Because the new subject--population gap is smaller than the population--image gap on every decoder, the population stage helps everywhere; how the gain divides between the two stages follows the population--image gap. On SAMGA, where this gap is smallest and the predicted anchors are closest to the images, the population stage alone is on par with direct calibration and the gain comes from the better start of the image stage ($+7.00$ in that stage, $+7.20$ overall). On ATM-S, where the gap is largest and the EEG embedding is low-dimensional (participation ratio 8.6 against 68.2 for its images), aligning directly to the images lowers Top-1 from 15.35 to 10.00\%, and even a labeled rotation toward the images reaches only 1.8\%, against 18.5\% toward the predicted anchors; the relay's $+7.65$ points over direct calibration come from the population stage. On BHA and Brain-HiVE both stages contribute.

\paragraph{A source-trained head for ATM-S.}
The ATM-S result points to its narrow EEG embedding, which a small change on the encoder side widens. We add a linear head on top of the frozen ATM-S encoder, initialized at the identity and trained with InfoNCE toward the frozen image features on source-training data only; early stopping and the weight penalty are chosen on source validation. The head raises the participation ratio of the target embedding from 8.6 to 14.2 and the CKA between population and images from .28 to .37, and the image stage no longer lowers accuracy ($-1.70$ to $+0.10$ points). The relay rises from 17.65 to 21.75\% and direct calibration from 10.00 to 13.10\%, so the relay stays 8.65 points ahead, for all ten subjects.

\section{Image reconstruction}
\label{app:reconstruction}
\begin{figure*}[t]\centering
\includegraphics[width=.86\textwidth]{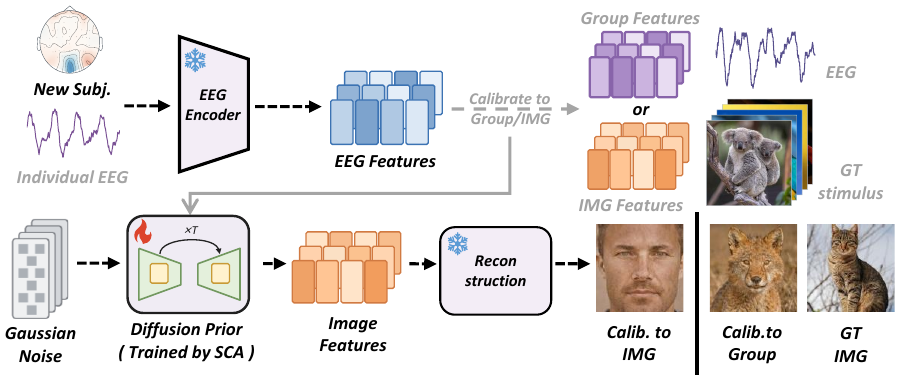}
\caption{\textbf{Reconstruction pipeline} (BHA). The frozen EEG encoder embeds the new subject's responses, and a single calibration aligns them to the measured population anchors of the training concepts (CHORUS; ``Group'' in the figure) or to their image embeddings (direct calibration; ``IMG'' in the figure). A diffusion prior trained with SCA maps the calibrated embedding to image features, which a frozen generator decodes. Bottom right: the held-out example S7 (cat) of \cref{fig:s_recon_matched1}.}
\label{fig:s_recon_pipeline}\end{figure*}
\begin{figure*}[t]\centering
\includegraphics[width=\textwidth]{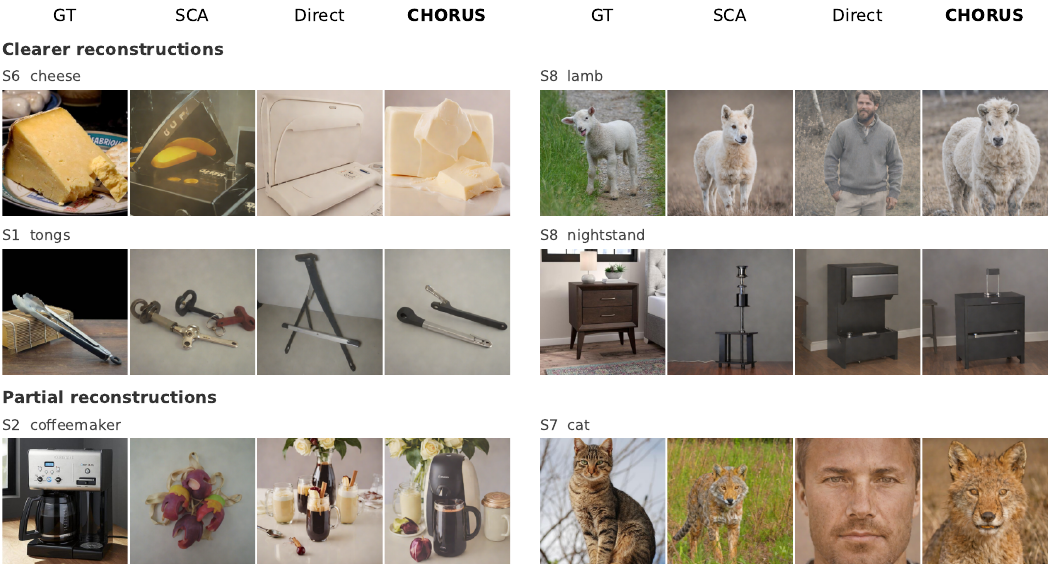}
\caption{\textbf{Population relay versus direct calibration} (BHA). Each block shares stimulus and generation indices. Top: clearer reconstructions; bottom: partial recovery. Direct calibration aligns to image prototypes of the training concepts; CHORUS aligns to their measured population prototypes.}
\label{fig:s_recon_matched1}\end{figure*}
\begin{table*}[t]
\centering\footnotesize\setlength{\tabcolsep}{4pt}
\caption{Reconstruction metrics (one training seed, ten subjects). Diffusion rows average two prior draws. Bold compares direct and population calibration on the same SCA pipeline. The oracle row uses source test EEG.}
\label{tab:s_recon}
\begin{tabular*}{\textwidth}{@{\extracolsep{\fill}}lrrrrrrrr@{}}\toprule
Condition & PixCorr$\uparrow$ & SSIM$\uparrow$ & Alex.2$\uparrow$ & Alex.5$\uparrow$ & Incep.$\uparrow$ & CLIP$\uparrow$ & Eff.$\downarrow$ & SwAV$\downarrow$\\\midrule
Base regression & 0.126 & 0.478 & 0.700 & 0.727 & 0.614 & 0.636 & 0.945 & 0.623\\
+ diffusion prior & 0.122 & 0.471 & 0.689 & 0.716 & 0.610 & 0.644 & 0.945 & 0.637\\
+ SCA & 0.141 & 0.457 & 0.716 & 0.753 & 0.638 & 0.682 & 0.932 & 0.627\\
Direct (training ref.) & \textbf{0.182} & \textbf{0.449} & 0.764 & 0.805 & 0.667 & 0.713 & 0.913 & 0.608\\
Population (training ref.) & 0.174 & 0.440 & \textbf{0.769} & \textbf{0.815} & \textbf{0.684} & \textbf{0.720} & \textbf{0.911} & \textbf{0.606}\\
\midrule
Group input (oracle) & 0.281 & 0.460 & 0.895 & 0.943 & 0.836 & 0.880 & 0.809 & 0.506\\
\bottomrule\end{tabular*}
\end{table*}

\begin{table*}[t]
\centering\footnotesize\setlength{\tabcolsep}{4pt}
\caption{Two-way CLIP reconstruction identification by held-out subject, with conditions as in Table~\ref{tab:s_recon}. Values average prior draws and generated samples.}
\label{tab:s_recon_sub}
\begin{tabular*}{\textwidth}{@{\extracolsep{\fill}}lrrrrrr@{}}\toprule
Subject & Base & Prior & SCA & Direct & Population & Group input\\\midrule
S1 & 0.622 & 0.635 & 0.653 & 0.713 & 0.726 & 0.883\\
S2 & 0.634 & 0.646 & 0.693 & 0.723 & 0.726 & 0.882\\
S3 & 0.578 & 0.617 & 0.638 & 0.711 & 0.724 & 0.878\\
S4 & 0.646 & 0.628 & 0.692 & 0.704 & 0.710 & 0.878\\
S5 & 0.606 & 0.637 & 0.676 & 0.711 & 0.715 & 0.885\\
S6 & 0.666 & 0.649 & 0.696 & 0.715 & 0.731 & 0.885\\
S7 & 0.648 & 0.666 & 0.685 & 0.682 & 0.698 & 0.876\\
S8 & 0.619 & 0.641 & 0.678 & 0.697 & 0.693 & 0.874\\
S9 & 0.639 & 0.645 & 0.667 & 0.681 & 0.687 & 0.881\\
S10 & 0.696 & 0.680 & 0.745 & 0.792 & 0.790 & 0.879\\
\bottomrule\end{tabular*}
\end{table*}

\subsection{Pipeline}
\Cref{fig:s_recon_pipeline} outlines the pipeline. Reconstruction uses the BHA encoders of one training seed in all ten folds and the same source split as retrieval. A high-level head predicts OpenCLIP ViT-H/14 embeddings with cosine and InfoNCE losses, followed by a diffusion prior~\cite{ramesh2022hierarchical,scotti2023reconstructing}; a low-level head predicts a VAE latent, and the target EEG also drives a spatial-conditioning branch from 17 occipito-parietal channels. The Base prior trains on nine views per image for 200 epochs and the SCA prior on 72 views for 25 epochs, the same number of updates; for each source, the 72 views comprise its own response, six pair averages and one quadruple average. Images are generated by SDXL-Turbo~\cite{sauer2024adversarial} with an IP-Adapter~\cite{ye2023ipadapter} (scale 1.5, three of five denoising steps); the prior uses 50 sampling steps and guidance two. Every condition uses two prior draws and three samples per draw, all retained.

\subsection{Calibration without test images}
Generation must not see the candidate test images, so reconstruction uses a single calibration stage toward training references. For each source-training concept we form a measured population prototype (the mean of source embeddings) and an image prototype (the mean of image embeddings). The 200 unlabeled test responses are matched to these prototypes by transport and centered orthogonal recovery, toward population prototypes for the relay and image prototypes for direct calibration. One oracle condition uses source test EEG: \emph{Group input} feeds the source-average test embedding to the prior and low-level head; it excludes the target subject and keeps the target's spatial branch.

\subsection{Results}
PixCorr and SSIM measure pixel and structural similarity; AlexNet layers 2 and 5, InceptionV3 and CLIP ViT-L/14 give two-way identification, and EfficientNet-B1 and SwAV ResNet-50 give feature distances (lower is better). Scores average samples, draws and stimuli within each subject, then subjects. Relative to direct calibration, the relay improves all four identification metrics (CLIP .720 versus .713) and both feature distances, but lowers PixCorr and SSIM (\cref{tab:s_recon}). Group input (.880) stays above the deployable relay for all ten subjects (\cref{tab:s_recon_sub}). Identification against the Group-generated images, using OpenCLIP features of the first sample of each draw, rises from 81.59\% with SCA alone to 86.12\% with direct calibration and 87.90\% with the relay: the relay pulls reconstructions toward those of the population. \Cref{fig:s_recon_matched1} shows matched examples with prior draw and sample zero.

\FloatBarrier
\section{Population-relay calibration on fMRI}
\label{app:fmri}
\subsection{Population geometry in native voxel space}
Does the population provide a closer reference even before training a shared decoder? In the Natural Scenes Dataset (NSD)~\cite{allen2022nsd}, we use 515 images presented three times to each of eight subjects. In each region, Pearson correlations between voxel patterns define an image-by-image representational similarity matrix (RSM). We average source RSMs in stimulus space, so voxels need not correspond across people. In each leave-one-subject-out fold, we compare the target's three-trial-mean RSM with each reference by Spearman correlation of the upper triangles (RSA). Stimulus identities are used in this offline analysis. The image feature is selected by RSA on source-private validation images.

The seven-source reference agrees more closely with the target than the source-selected visual features in V1, the parahippocampal place area (PPA) and ventral visual cortex (\cref{tab:s_fmri}a). Measurement budget matters: with three measurements per image, averaging three single-trial RSMs across people yields higher RSA than averaging those RSMs within one person in every region. Averaging one person's three voxel patterns \emph{before} computing their RSM is stronger than either. Thus the full population captures shared stimulus structure, while the benefit of pooling across people depends on how repeats are combined.

\subsection{Changing the first-stage reference}
We use the official one-session MindEye2 checkpoints~\cite{scotti2024mindeye2} for subjects 01, 02, 05 and 07. Each checkpoint has already seen one training session from its target subject. Subsequent calibration uses an unlabeled batch of target responses and candidate images. For each of the seven other subjects, we fit an image-to-response ridge predictor on that source's private training images, select hyperparameters on source validation data, and average the seven predictions for each candidate. Predictor fitting and selection exclude the 1,000 shared test images. On these images the relay improves over direct calibration for every target (\cref{tab:s_fmri}b).

\begin{table}[!htb]
\centering
\caption{\textbf{Population geometry and fMRI retrieval on NSD.} (a) Mean RSA across eight held-out subjects. (b) Top-1 (\%) for the four one-session MindEye2 checkpoints; the mean weights subjects equally. Only the measured relay uses source responses to test images.}
\label{tab:s_fmri}
\footnotesize
\begin{tabular*}{\columnwidth}{@{\extracolsep{\fill}}lrrr@{}}\toprule
\multicolumn{4}{@{}l}{\textbf{(a) Native voxel geometry, 515 images}}\\
Reference & V1 & PPA & Ventral\\\midrule
Source-selected image features & 0.163 & 0.246 & 0.243\\
One source, three trials (average betas) & 0.369 & 0.431 & 0.566\\
One source, three trials (average RSMs) & 0.329 & 0.361 & 0.504\\
Three sources, one trial each & 0.342 & 0.368 & 0.519\\
Seven sources, three trials each & 0.537 & 0.601 & 0.717\\
\bottomrule\end{tabular*}\par\smallskip
\setlength{\tabcolsep}{1.3pt}
\begin{tabular*}{\columnwidth}{@{\extracolsep{\fill}}lrrrrr@{}}\toprule
\multicolumn{6}{@{}l}{\textbf{(b) Retrieval from frozen fMRI decoders}}\\
Reference / route & 01 & 02 & 05 & 07 & Mean\\\midrule
\multicolumn{6}{@{}l}{\emph{1,000-way retrieval}}\\
Native + CSLS & 86.00 & 79.80 & 50.30 & 47.00 & 65.78\\
Direct calibration & 84.60 & 77.30 & 48.20 & 47.10 & 64.30\\
Predicted population relay & 99.70 & 98.80 & 70.90 & 69.10 & \textbf{84.63}\\
\addlinespace[2pt]
\multicolumn{6}{@{}l}{\emph{907-way retrieval, same two-stage schedule}}\\
Image $\to$ image & 90.19 & 81.48 & 54.69 & 55.68 & 70.51\\
Predicted population $\to$ image & 99.89 & 98.79 & 74.75 & 74.31 & \textbf{86.93}\\
Measured population $\to$ image (oracle) & 99.78 & 98.90 & 76.41 & 78.28 & 88.34\\
\bottomrule\end{tabular*}
\end{table}

\paragraph{Same two stages, different first reference.}
For the 907 images measured in all seven sources, we keep the queries, transport initialization procedure, two-stage matching and refitting schedule, final image gallery and CSLS readout fixed. Only the first reference changes, including its centering mean: image embeddings with the gallery mean, or predicted anchors with the source population mean. First-stage pair accuracy rises from 70.53\% to 84.67\%. The image stage thus starts from better correspondences and final retrieval improves from 70.51\% to 86.93\%, a gain of 16.43 points across four subjects. For comparison, using the measured population as an offline oracle gives 88.34\%.

\FloatBarrier

\end{document}